\pdfoutput=1
\documentclass[preprint,aps,prd,nofootinbib]{revtex4-1}
\usepackage{amssymb,amsmath,graphicx,color,microtype,enumerate,slashed,hyperref,xcolor,cases,verbatim}

\hypersetup{colorlinks=true,linkcolor=blue,filecolor=blue,urlcolor=blue, citecolor=blue} 

\begin{document}
\title{Gravitational Production of Superheavy Dark Matter and Associated Cosmological Signatures} 
\author{Lingfeng Li$^{1}$, Tomohiro Nakama$^{1}$, Chon Man Sou$^{2,1}$, Yi Wang$^{2,1}$, Siyi Zhou$^{2,1}$}  
\email{iaslfli@ust.hk,iasnakama@ust.hk,cmsou@connect.ust.hk, phyw@ust.hk, szhouah@connect.ust.hk}
\affiliation{${}^1$Jockey Club Institute for Advanced Study, The Hong Kong University of Science and Technology, \\
	Clear Water Bay, Kowloon, Hong Kong, P.R.China} 
\affiliation{${}^2$Department of Physics, The Hong Kong University of Science and Technology, \\
	Clear Water Bay, Kowloon, Hong Kong, P.R.China \vspace{0.03\textheight}}

\begin{abstract}
	We study the gravitational production of super-Hubble-mass dark matter in the very early universe. 
	We first review the simplest scenario where dark matter is produced mainly during slow roll inflation. Then we move on to consider the cases where dark matter is produced during the transition period between inflation and the subsequent cosmological evolution. The limits of smooth and sudden transitions are studied, respectively. The relic abundances and the cosmological collider signals are calculated.
\end{abstract} 
\maketitle

\section{Introduction}
$\Lambda$CDM cosmology has succeeded in explaining the existence and structure of the cosmic microwave background (CMB),
the large-scale structure (LSS), the abundances of the light elements and the accelerating expansion of the universe. However, the nature of cold dark matter
(CDM) \cite{Bertone:2004pz}, including its production mechanism and interactions, is still not known.  

There are many production mechanisms of dark matter, one of which is gravitational particle production \cite{Ford:1986sy}. In this case, dark matter can be produced during inflation and in post-inflationary phases as well. The relevant models of dark matter include Planckian Interacting Dark Matter (PIDM) \cite{Garny:2015sjg,Garny:2017kha,Hashiba:2018iff,Haro:2018zdb,Hashiba:2018tbu}, WIMPZILLA \cite{Kolb:1998ki,Kolb:2007vd}, SUPERWIMP \cite{Feng:2010gw}, FIMP \cite{Hall:2009bx} (the model considered in \cite{Garny:2015sjg} can be viewed as ``FIMPZILLA"). 

In this paper, we focus on Superheavy Dark Matter (SHDM) \cite{Chung:1998zb,Kuzmin:1998uv,Chung:1999ve,Kuzmin:1999zk,Chung:2001cb}, for which $m_X> H$. The existence of such dark matter may originate from supersymmetry breaking theories \cite{Delacretaz:2016nhw}, string inspired models \cite{Chang:1996rf,Chang:1996vw,Faraggi:1999iu,Coriano:2001mg} or Kaluza-Klein theory of extra dimension \cite{Garny:2015sjg,Garny:2017kha,Kumar:2018jxz}. The amount of gravitational dark matter production during inflation is well known and it is roughly proportional to $H^3\mu^3e^{-2\pi\mu}$, where $\mu\equiv\sqrt{m_X^2/H^2-9/4}$. In reality, after inflation, there will eventually be the radiation-dominated universe. There are a variety of mechanisms of the so-called preheating/reheating period sandwiched between the inflation and radiation-dominated universe \cite{Kofman:1994rk,Kofman:1997yn} (see \cite{Bassett:2005xm,Boyanovsky:1996sv,Allahverdi:2010xz,Frolov:2010sz,Amin:2014eta} for reviews). When the mass of the gravitationally produced dark matter particle is large, the production rate is in general exponentially suppressed in terms of the mass of the dark matter particle. In this case, the produced superheavy dark matter particle may not be sufficient to explain the existing dark matter relic abundance. However, as noted recently in \cite{Ema:2018ucl}, during an inflaton oscillation regime \cite{Ema:2015dka,Ema:2016hlw}, where the scale factor oscillates very rapidly, dark matter particles with number density proportional to $H^3$ can be produced. An analytical approach for this case is developed recently in \cite{Chung:2018ayg}. 

We discuss that there is another scenario where the number density of order $H^3$ superheavy dark matter can be produced, where there is a sharp transition between the inflation-radiation period. This may be realized in some inflationary models such as quintessential inflation \cite{Peebles:1998qn}, as discussed later. 

%W When the transition timescale is sufficiently long, the particle production is often subject to exponential suppression, as we will demonstrate using a toy universe model of inflation-Minkowski transition.  

%In this paper, we would like to introduce some novel concepts and techniques into the analysis of gravitationally produced dark matter. One novel technique we would like to

We will also revisit smooth transition cases by introducing the Stokes line method, which was used in the analysis of the Schwinger effect \cite{Dabrowski2014}. In the community of the gravitational dark matter production, the common method is to calculate the Bogoliubov coefficients in the WKB approximation \cite{Chung:1998bt,Quintin:2014oea,Celani:2016cwm} which assumes that the mode functions of the fields are close to adiabatic mode functions. Quantum transitions such as excitations with slowly changing Hamiltonian  \cite{Berry1990,Berry1993,Betz2004}
and particle productions in slowly changing background \cite{Winitzki:2005rw} are exponentially small, implying that the desired Bogoliubov coefficients can not be calculated by introducing a normal series solution with integer orders. On the other hand, the characteristics of mode functions change 
considerably during the universe's expansion, and we have to identify carefully the dominant and subdominant parts in the WKB ansatz \cite{Dumlu2010} in order to interpret the Bogoliubov coefficients as particle productions. Some researchers \cite{Kim2010,Kim2013,Dumlu2010,Dabrowski2014,Dabrowski2016}
considered these questions and argued that the evaluations of particle
productions require more information besides WKB ansatz, i.e. the
Stokes phenomenon which involves the emergence of subdominant component with negative frequency \cite{Higham2015}. These studies imply that the Stokes phenomenon can indicate the production events and provides results with a reasonable precision, and therefore we may adopt this interpretation. On the other hand, the ref.~\cite{Berry1989} showed that the superadiabatic approximation can describe universal and smooth particle productions when the time evolution hits the Stokes line. This method utilizes Dingle's theory of asymptotic series \cite{Dingle1973}. We adopt this method to calculate the gravitational dark matter production in a smoothly changing background. 

There are many models for gravitationally produced dark matter with different mass ranges and different types of interactions \cite{Chung:2004nh,Chung:2011ck,Chung:2013rda,Kannike:2016jfs,Alonso-Alvarez:2018tus,Kolb:2017jvz,Fairbairn:2018bsw,Garny:2018grs}. One may ask if it is possible to determine some properties of such dark matter independently of the details of these models. One such possibility is to make use of the cosmological-collider signals \cite{Chen:2009we,Chen:2009zp,Baumann:2011nk,Noumi:2012vr,Arkani-Hamed:2015bza}. By measuring the squeezed-limit non-Gaussianity, from the frequency of its oscillation, we can in principle read off the mass of the dark matter particle. The spin information may also be read off from the angular dependence in the squeezed limit. Note though that the frequency depends on the inflationary Hubble scale and possible non-minimal gravitational couplings as well. Also, the coupling between dark matter and the inflaton may affect the relic abundance of dark matter, or correct the mass of the dark matter particle in a significant way, as we will discuss later.

This paper is organized as follows: In Section~\ref{relic}, we setup the model and study three scenarios of the dark matter gravitational production in the early universe. The dark matter relic abundance is also calculated. In Section~\ref{collider}, we discuss associated cosmological collider signals which may help to determine the mass of the gravitationally produced super-heavy dark matter particle. We conclude in Section~\ref{summary}.

\section{Dark Matter and its Relic Abundance} \label{relic}

Consider a FRW universe with the metric $ds^2 = - d t^2 + a^2(t) d \mathbf x^2$. We denote the dark matter field as $X$. The action is
\begin{align}\label{actionDM}
	S_X = - \int d^4 x \sqrt{-g} \bigg[\frac{1}{2} (\partial_\mu X)^2 + \frac{1}{2} m_X^2 X^2 + \frac{1}{2} \xi R X^2 \bigg]~,
\end{align}
where $R$ is the Ricci scalar, and the last term represents the non-minimal coupling. 
%Note that $\xi = 1/6$ is the conformally-coupled case. 

The equation of motion for $X$ is
\begin{align}
	\ddot X + 3 H \dot X - \frac{1}{a^2} \nabla^2 X + M_X^2 X =0, \quad M_X^2 \equiv m_X^2 + \xi R ~.
\end{align}
We can write the field $X$ in terms of modes as
\begin{align}
	X = \int \frac{d^3 \mathbf k}{(2\pi)^3} e^{i \mathbf k\cdot \mathbf x} a^{-3/2} [f_k a_{\mathbf k} + f_k^* a_{-\mathbf k}^\dagger ]~,
\end{align}
where $a_\mathbf k$ and $a_{-\mathbf k}^\dagger$ are the annihilation and creation operators that satisfy the commutation relations $[a_{\mathbf k}, a_{\mathbf k'}] = 0$ and $[a_{\mathbf k}, a_{\mathbf k'}^\dagger] = (2\pi)^3 \delta^{(3)} (\mathbf k - \mathbf k')$. We find
\begin{align}
	 \ddot f_k(t)  + \omega_k^2 f_k (t) = 0, \quad \omega_k^2=  \frac{k^2}{a^2} + H^2 \mu^2 - \frac{3}{2} \dot H , \quad \mu\equiv \sqrt{\frac{m_X^2}{H^2}+12 \xi-\frac{9}{4}} ~,
\end{align}
where $H$ is the Hubble parameter here. Note that $\dot H$ is very small in the slow-roll inflation cases and vanishes in the case of exact de Sitter space. 

We will consider three scenarios of the gravitational production of superheavy dark matter. The production in an exact de Sitter phase is reviewed in Section~\ref{eternalinflation}. We use the Stokes 
line method to estimate the particle production in a toy universe in which inflation is connected to Minkowski spacetime in Section~\ref{inflationMinkowski}. Section~\ref{suddentransition} is devoted to an analysis of a sudden transition between the inflation and radiation period.  
%{\color{blue}in this section, it would be better to use $H_e$ instead of $H$ as long as $\dot{a}/a$ is not a constant.}
\subsection{Slow Roll Inflation}\label{eternalinflation}
In this section, we review the gravitational particle production during slow roll inflation, approximated by a period of de Sitter expansion. This part is well known and recently reviewed in \cite{Markkanen:2016aes}. The particle production in the presence of a background electric field is discussed in \cite{Kobayashi:2014zza}. 

During slow roll inflation, the scale factor is approximately $a(t) = e^{Ht} = -1/(H \tau)$, and $R = 12 H^2$. When inflation ends, the scale factor starts to evolve in a non-accelerated way. There are two classes of solutions to the massive field equation of motion, corresponding to ``in" state and ``out" state, respectively \cite{Chen:2009we},
\begin{align}\label{instatemodefunction}
	f_k^{\rm in} (t) & = \sqrt{\frac{\pi}{4H}} e^{-\pi\mu/2} H_{i\mu}^{(1)} (-k\tau)~, \\ \label{outstatemodefunction}
	f_k^{\rm out} (t) & = \bigg(\frac{2H}{k}\bigg)^{i\mu} \frac{\Gamma(1+i\mu)}{\sqrt{2H\mu}} J_{i\mu} (-k\tau)~,
\end{align}
where $H^{(1)}_{\nu}(x)$ is the Hankel function of the first kind, and $J_{\nu}(x)$ is the Bessel function. 

The mode functions are related via a Bogoliubov transformation as
\begin{align}\label{eq:expand_Bogoliubov}
	f_k^{\rm in} (t) = \alpha_k f_k^{\rm out} (t) + \beta_k f_k^{\rm out *} (t) ~.
\end{align}
Inserting the explicit expressions for the ``in" state mode function and ``out" state mode function, we obtain the following expressions for the Bogoliubov coefficients
\begin{align}
	\beta_k = \bigg(\frac{2H}{k}\bigg)^{i\mu} \frac{e^{\pi\mu/2}\sqrt{2\pi\mu}}{(1-e^{2\pi\mu})\Gamma[1-i\mu]}~, \quad \alpha_k = - e^{\pi\mu} \beta_k^*~,  
\end{align}
then $|\beta_k|^2$ can be evaluated as
\begin{align}
	|\beta_k|^2 = \frac{1}{e^{2\pi\mu}-1}~.
\end{align}  
Then 
\begin{align}\label{num_density1}
	N_X = \int_{0}^{\infty}  dk \,\, 2\pi  k^2 |\beta_k|^2~.
\end{align}
gives the total number of particles produced per comoving three-volume, from the past infinity to future infinity, as shown below (see also \cite{Kobayashi:2014zza}). We can focus on the time duration when $|\omega_k'/\omega_k^2|$ is largest. By studying the maximum value of $|\omega_k'/\omega_k^2|$, we know that particles are produced mostly around the time
\begin{align}
	- k \tau \sim \mu ~.
\end{align}
Using this relation to rewrite the $k$ integral into $\tau$ integral, we have
\begin{align}
	\int_0^\infty dk\,\, k^2 = \mu^3 \int_{-\infty}^0 d\tau\,\, \bigg(-\frac{1}{\tau}\bigg)^4 = \mu^3 \int_{-\infty}^0 d\tau\,\,  (aH)^4~.
\end{align}
Then the integral in \eqref{num_density1} can be evaluated as
\begin{align}\label{num_density}
N_X = 2\pi |\beta_k|^2 \mu^3 \int_{-\infty}^{0} d\tau\,\,  (aH)^4 ~.
\end{align}
The production rate per unit physical four-volume can be obtained as
\begin{align}
	\Gamma =  \frac{H^4 \mu^3 2\pi}{e^{2\pi\mu}-1}~.
\end{align}
The physical number density of particles at each moment can be evaluated as:
\begin{align}
	n_X  =  \frac{1}{a(\tau)^3} \int_{-\infty}^{\tau} d\tilde \tau a (\tilde \tau)^4 \Gamma = \frac{\Gamma}{3 H} ~.
\end{align}
which is constant in time. Hence, as a particular case, $n_X$ can be regarded as the number density at the end of inflation. Then the current DM abundance is~\cite{Chung:2001cb}:
\begin{equation}
\Omega_X h^2 \simeq \frac{8\pi}{3}\Omega_{\text{R}}h^2 \frac{m_X n_X}{M_{\rm pl}^2 H^2}\bigg( \frac{T_{\rm RH}}{T_0} \bigg)\simeq 1.14\times 10^{9} \mu^3 e^{-2\pi\mu} \frac{H m_X T_{\rm RH}}{M_{\rm pl}^2}
\end{equation}
in the unit of GeV. Here $\Omega_{\rm R}$ and $T_0$ are the radiation energy fraction and temperature today. This shows that the relic density is exponentially sensitive to $\mu$. If we further assume that reheating happens soon enough after inflation ends, $T_{\rm RH}\simeq \sqrt{H M_{\rm pl}/g_R}$ by energy conservation and $\mu=1$, we will need $H \gtrsim 10^9$~GeV in order to make $X$ a major component of DM. Here we assume that production during the exact de Sitter phase is dominant and determines the eventual relic density, but it may be subdominant depending on post-inflationary phases, as we will discuss below.

\subsection{Inflation Connected to Minkowski Spacetime}\label{inflationMinkowski}
In this section, we discuss the case where inflation is  smoothly  connected to Minkowski spacetime. The particle production in this scenario would approximate the particle production in more realistic scenarios where inflation is connected to a stage of the universe with much lower Hubble scale (and thus approximately Minkowski), for example, radiation-dominated universe with transition time of order $H^{-1}$.   

We introduce the Stokes line method to compute the dark matter relic abundance. The method utilizes the fact that the adiabatic expansion from the WKB approximation is a factorially divergent asymptotic series which can be resummed by the Borel's summation. Under a proper truncation of the asymptotic series, smooth particle production is obtained while the physical time is crossing the Stokes line of mode functions \cite{Berry1989}.

We parametrize our toy model of the scale factor evolution as
\begin{align}
a(t) = \frac{e^{H t}}{1+ e^{H t}}~.
\end{align}
Note that in this subsection, $H$ is the initial Hubble parameter, which is a constant and not the Hubble parameter $\dot a/a$ for all times.

For $t\ll 0$, the scale factor $a(t)$ approaches de Sitter spacetime. For $t\gg 0$, the scale factor $a(t)$ approaches Minkowski spacetime with $a$ approaching unity. We would like to use the Stokes line method to calculate the particle production in this toy model. This study would shed some light on the particle production in the real universe. We present the main steps in the following and leave a detailed review of this method in Appendix~\ref{sec:app-div}. Applications of this method to the inflationary universe is also given in this Appendix~\ref{sec:app-div}.

We use $t_c$ to denote the complex time on lower half-plane satisfying $\omega_k(t_c) = 0$, 
%To see where those zero points are, we plot the real and imaginary part of $\omega_k$ in FIG~\ref{REANDIMAGINEOFW}.
and $f_k(t)$ can be expressed as
\begin{align}
f_{k}(t)  =\frac{\mathrm{exp}(-i\int_{t_{i}}^{t}\omega_{k}dt' )-iS_{k}(t)\mathrm{exp}(i\int_{t_{c}}^{t_{c}^{*}}\omega_{k}dt' )\mathrm{exp}(i\int_{t_{i}}^{t}\omega_{k}dt' +i\phi )}{\sqrt{2\omega_{k}}}\ ,\label{eq:approx}
\end{align}
where $S_k$ is the Stokes multiplier which indicates the moment when the Stokes line is hit, and $\phi$ is an unimportant constant phase. Thus, the Bogoliubov coefficients are
\begin{align}
\alpha_{k}(t)\approx1,~
\beta_{k}(t)\approx -iS_{k}(t)\mathrm{exp}\Big(i\int_{t_{c}}^{t_{c}^{*}}\omega_{k}dt' \Big)~.\label{eq:Bogoliuaa}
\end{align}

\begin{figure}[htbp] 
	\centering 
	\includegraphics[width=8cm]{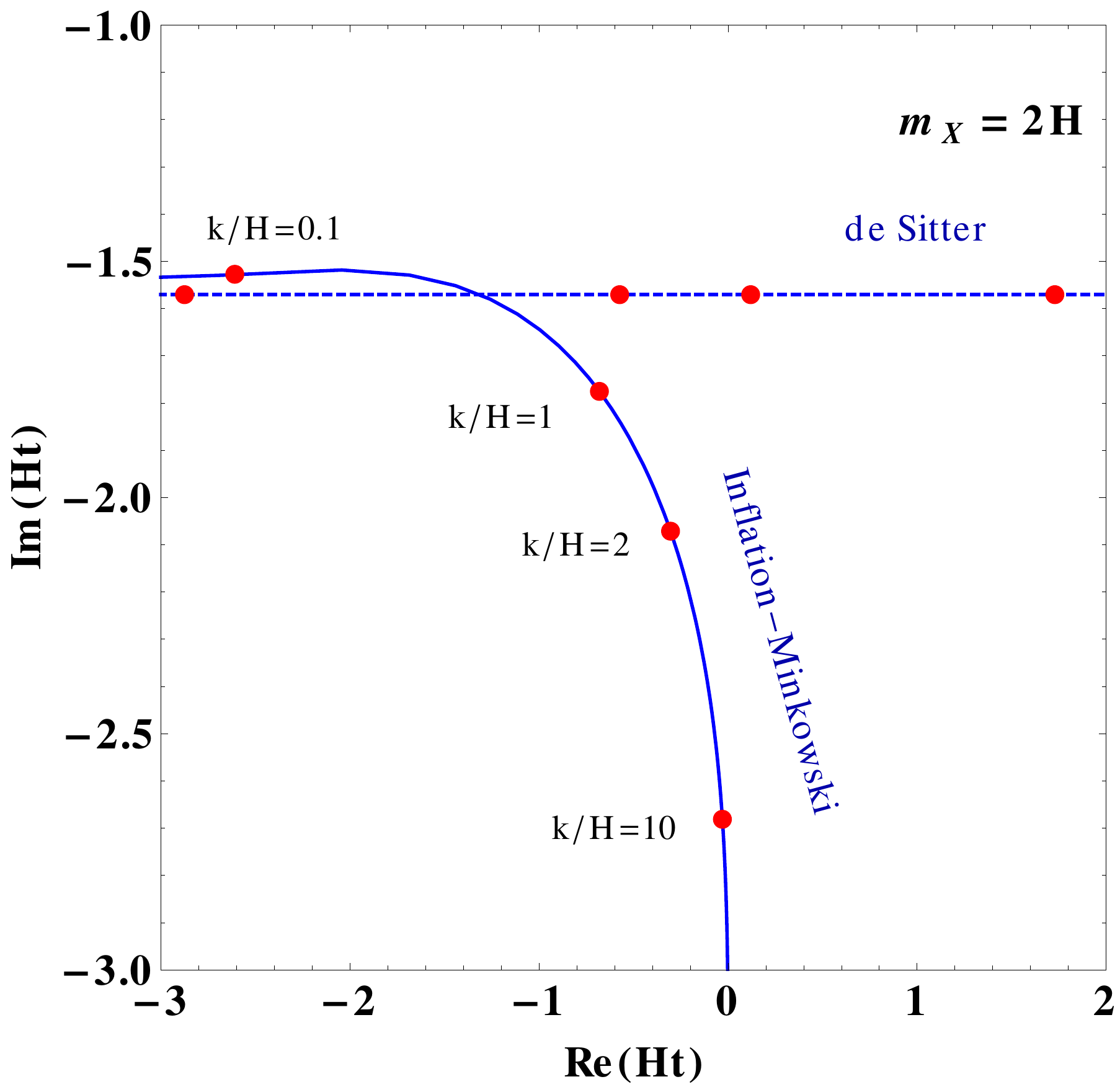}  
	\includegraphics[width=7.7cm]{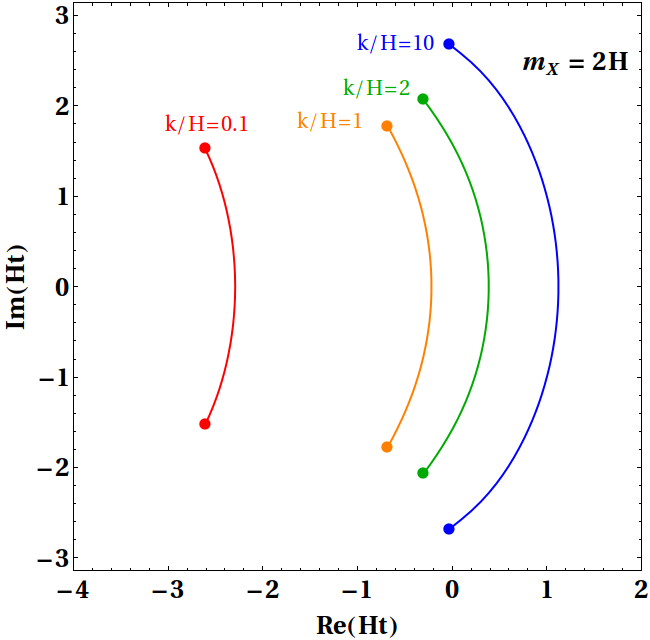}  
	\caption{{\bf Left: }$t_c$ with different $k$. The solid blue curve corresponds to the trajectory of $t_c(k)$ in our toy model of smooth transition. The dashed blue curve is the corresponding trajectory for de Sitter. 
	%with $m_X=2H$ and $k/(a_f H)$=0.1,~1,~2 and 10. 
%Due to the conformal invariance, de Sitter solutions of different $k$'s only differ by a translation in $t$. On the contrary, for subhorizon modes with $k/a_f \gtrsim H$, $t_c$ deviates from de Sitter solution, lead to a large exponential suppression in $\beta_k$ from Eq.~(\ref{eq:Bogoliuaa}). 
{\bf Right: }The Stokes lines defined in Appendix~\ref{sec:app-div} with the same conditions as the left panel. Each Stokes line passes though $t_c$ and $t_c^*$.
%Their intersections with the real time axis are roughly the moments when each $k$ modes are produced, which can extend to $t>0$ regime.
}   \label{plotofe}
\end{figure}
\begin{figure}[htbp] 
	\centering 
	\includegraphics[width=8cm]{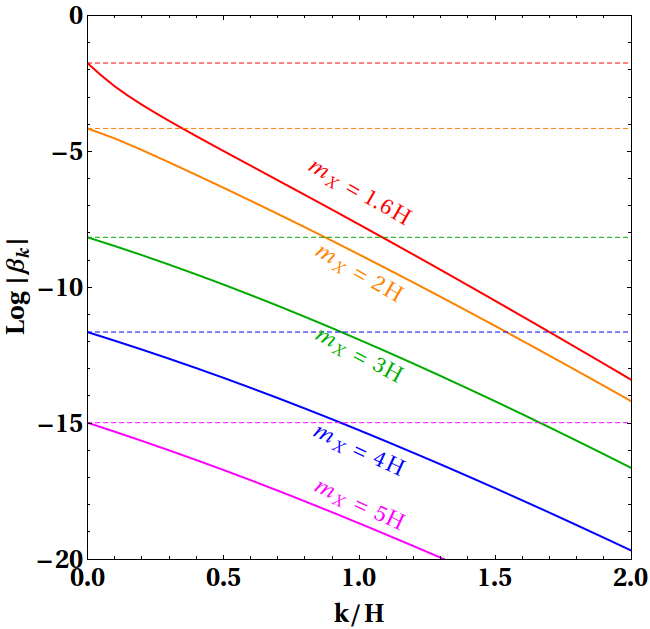}  
	\includegraphics[width=8.2cm]{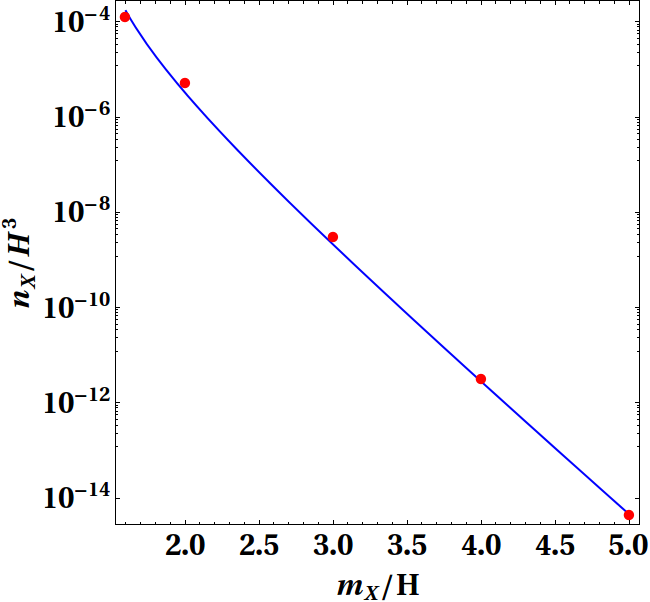}  
	\caption{\textbf{Left: }$\log |\beta_k|$ as a function of $k$ for different $m_X$. The exact de Sitter solutions correspond to the dashed lines (See Section~\ref{eternalinflation}) while the solid curves are evaluated from Eq.~(\ref{eq:Bogoliuaa}). \textbf{Right: }$n_X$ as a function of the mass of the dark matter $m_X$. The red dots are from numerical integration of Eq.~\eqref{eq:Bogoliuaa}. The blue line is the numerical fitting result from Eq.~\eqref{empirical}. }   \label{numberdensity}
\end{figure}

 
We draw the trajectory of $t_c$ solutions for both de Sitter and our inflation-Minkowski case on the left panel of FIG.~\ref{plotofe}. We also draw some of the Stokes lines for different $k$ modes on the right panel. We then integrate Eq.~\eqref{eq:Bogoliuaa} numerically. The results are present on the left panel of FIG.~\ref{numberdensity}. The numerical results can be fitted by 
\begin{align}\label{empirical}
n_X \simeq 10^{-2} H^3 \mu e^{-2\pi\mu}~.
\end{align}
The equation describes $n_X$ estimated by numerical integration well for a wide range of $m_X/H$, shown on the right panel of Fig.~\ref{numberdensity}. 

To be consistent with observations, we require
%\begin{align}
%\Omega_X h^2 & \simeq 4.31\times 10^{-5} \frac{T_{\rm RH}}{T_0} \frac{\rho_X(t_e)}{\rho_I(t_e)}   = 4.31\times 10^{-5} \frac{T_e}{T_0} \frac{ A H^3 \mu e^{-2\pi\mu} m_X }{3M_{\rm pl}^2 H^2} = 0.12 ~,
%\end{align}
\begin{align}
\Omega_X h^2 & \simeq\frac{8\pi}{3}\Omega_{\text{R}}h^2 \frac{m_X n_X}{M_{\rm pl}^2 H^2}\bigg( \frac{T_{\rm RH}}{T_0} \bigg)\simeq  4.31\times 10^{-7} \frac{T_{\rm RH}}{T_0} \frac{  H \mu  m_X }{3M_{\rm pl}^2}e^{-2\pi\mu} = 0.12 ~.
\label{eq:OmegaDMStokes}
\end{align}
%If we further assume that reheating happens soon enough after inflation ends, $T_e=\sqrt{H M_{\rm pl}}$, $T_0\simeq 10^{-3}{\rm eV}$.  

\begin{figure}[th]
\includegraphics[scale=0.5]{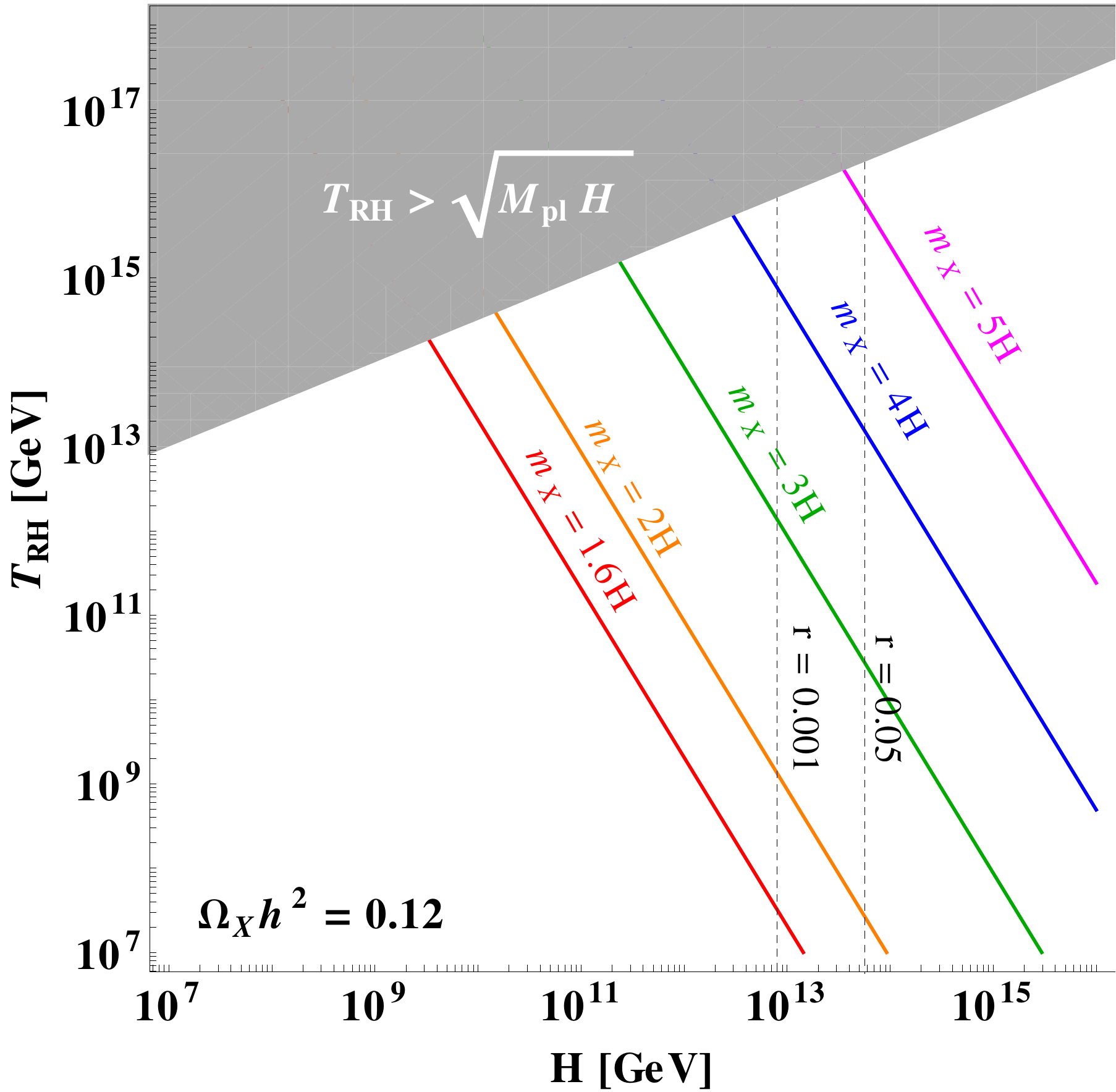}
\caption{The parameter space that inflation can produce enough DM according to Eq.~\ref{eq:OmegaDMStokes}. The gray region is forbidden by energy conservation.}
\label{fig:Hscan}
\end{figure}

In single-field slow-roll models of inflation, the initial Hubble parameter can be related to the tensor-to-scalar ratio $r$ as \cite{Enqvist:2017kzh}
\begin{align}
	H = 8\times 10^{13} \sqrt{r/0.1} {\rm GeV}~. 
\end{align} 
For $r=0.05$, we have $m_X/H\sim 4.93$ to explain the full dark matter relic abundance. The corresponding value for $r=0.001$ is $m_X/H \sim 4.125$. We draw the parameter space compatible with $\Omega_{\rm DM}=\Omega_X$ in Figure.~\ref{fig:Hscan}.

%\begin{figure}[htbp] 
%	\centering 
%	\includegraphics[width=8cm]{beta.png}  
%	\caption{$\log |\beta_k|$ as a function of $k$ for different $m_X$. The exact de Sitter solutions correspond to the dashed lines (See Section~\ref{eternalinflation}) while the solid curves are evaluated from Eq.~(\ref{eq:Bogoliuaa}).}   \label{betakkk}
%\end{figure} 
%
%\begin{figure}[htbp] 
%	\centering 
%	\includegraphics[width=8cm]{numerical.png}  
%	\caption{$n_X$ as a function of the mass of the dark matter $m_X$. The red dots are from numerical integration of Eq.~\eqref{eq:Bogoliuaa}. The blue line is the numerical fitting result from Eq.~\eqref{empirical}. }   \label{numberdensity}
%\end{figure} 

\subsection{Sudden Transition between Inflation and Radiation Domination}\label{suddentransition}
In this subsection we aim at evaluating the particle production during a sudden inflation-radiation transition, with the transition time scale $\Delta t\ll H^{-1}$. Such a situation may be realized by modifying the inflaton potential for quintessential inflation \cite{Peebles:1998qn}. Radiation can also be produced gravitationally, or one can also introduce coupling between the inflaton and other fields to reheat the universe \cite{Dimopoulos:2017tud,Hashiba:2018tbu}. See also \cite{Nakama:2018gll}. In this case, we can parametrize the evolution history of the universe $a(\tau)$ in the following way, 
\begin{numcases}{a(\tau)=} 
	\displaystyle -\frac1{H\tau}~,  \quad   {\rm when}~\tau<\tau_{\rm end}~.\label{equation622} \\ \label{equation623}
	\displaystyle \frac{1}{H \tau_{\rm end}^2} (\tau - 2 \tau_{\rm end})  ~, \quad  {\rm when}~ \tau>\tau_{\rm end}~;
\end{numcases}
such that $a$ and $\dot a$ are continuous. Note that in this subsection, $H$ is a constant parameter (initial Hubble parameter in the inflation stage), which is no longer interpreted as the Hubble parameter $\dot a/a$ after the sudden transition.

The dark matter $X$ is quantized as 
\begin{align}
X_{\mathbf k} (t) = u_k (t) a_{\mathbf k} + u^*_k (t) a^\dagger_{-\mathbf k}~.
\end{align}
Let us focus on subhorizon modes with $k_{\rm min} \equiv a m_X <k$, for which the mass term can be neglected. We also assume $\xi = 0$ for simplicity. The mode function satisfies the Mukhanov-Sasaki equation
\begin{align}
(a u_k)'' + \bigg(k^2 - \frac{a''}{a} \bigg) (a u_k) = 0.
\end{align}
For the modes with $k\ll a \delta t^{-1} \equiv k_{\rm max}$, the solutions can be approximated as follows. First, the above equation is solved by
\begin{numcases}{u_k(\tau)=} 
	\displaystyle \frac{H}{\sqrt{2 k^3}} (1+i k \tau) e^{-i k \tau}  , \,  {\rm when} \,\tau<\tau_{\rm end}~.\label{equation624} \\ \label{equation625}
	\displaystyle  \frac{c_1 H \tau_{\rm end}}{\sqrt{2 k} (\tau-2\tau_{\rm end})} e^{-i k(\tau-2\tau_{\rm end})} +  \frac{c_2 H \tau_{\rm end}}{\sqrt{2 k} (\tau-2\tau_{\rm end})} e^{i k(\tau-2\tau_{\rm end})}  , \, {\rm when} \,\tau>\tau_{\rm end}~;
\end{numcases}
the coefficients $c_{1,2}$ can be determined by the connecting conditions $u_k(\tau_{\rm end}^-) = u_k(\tau_{\rm end}^+)$ and $u'_k(\tau_{\rm end}^-) = u'_k(\tau_{\rm end}^+)$ as follows:
\begin{align}
c_1 = \frac{e^{ - 2 i k \tau_{\rm end} }  (i+2 k \tau_{\rm end}(-1-i k \tau_{\rm end}) ) }{2 k^2 \tau_{\rm end}}, \quad c_2 = -\frac{i}{2 k^2 \tau_{\rm end}}~.
\end{align}
\begin{figure}	
	\includegraphics[scale=0.7]{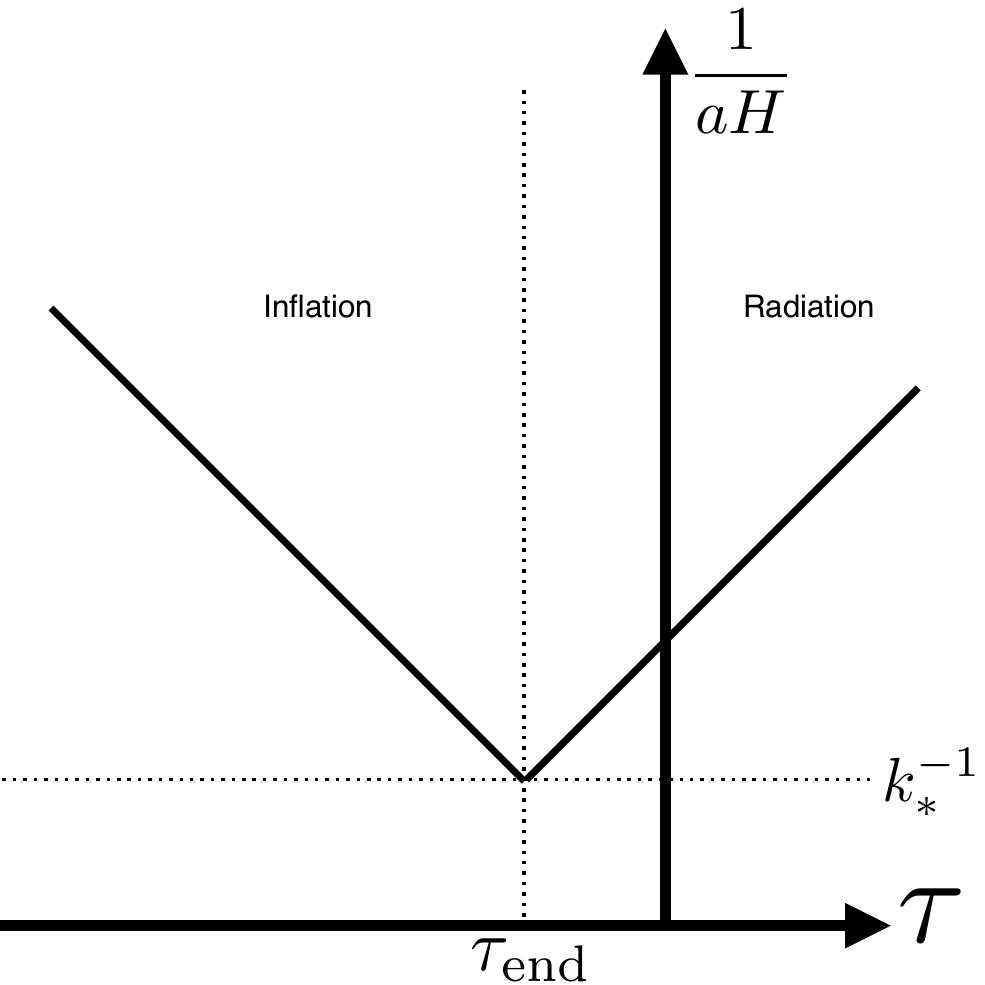}
	\caption{Sketch of the evolution of the comoving Hubble horizon in the sudden transition scenario.} 
\end{figure}
From these we find 
\begin{align}
|c_1|^2-|c_2|^2 = \tau_{\rm end}^2~.
\end{align}
Inserting the expression of $c_1$ and $c_2$ into the expression for $u_k(\tau)$ when $\tau>\tau_{\rm end}$,
\begin{align}\nonumber
u_k(\tau) & = \frac{e^{ - 2 i k \tau_{\rm end} }  (i+2 k \tau_{\rm end}(-1-i k \tau_{\rm end}) ) }{2 k^2 \tau_{\rm end}^2 } \frac{H \tau_{\rm end}^2 }{\sqrt{2 k} (\tau-2\tau_{\rm end})} e^{-i k(\tau-2\tau_{\rm end})} \\
& -\frac{i}{2 k^2 \tau_{\rm end}^2 } \frac{H \tau_{\rm end}^2 }{\sqrt{2 k} (\tau-2\tau_{\rm end})} e^{i k(\tau-2\tau_{\rm end})}  ~.
\end{align}
On the other hand, the mode function $u_k(\tau)$ can be written as
\begin{align}
u_k(\tau) = \alpha_k u_{\rm out} (\tau) + \beta_k u^*_{\rm out} (\tau)~,
\end{align}
where $u_{\rm out}$ is the positive-frequency component of the mode function with the normalization condition
\begin{align}
u_{\rm out} u'^{*}_{\rm out} - u'_{\rm out} u^*_{\rm out} = \frac{i}{a^2}~.
\end{align}
So the Bogoliubov coefficients are given by
\begin{align}
\alpha_k = \frac{e^{ - 2 i k \tau_{\rm end} }  (i+2 k \tau_{\rm end}(-1-i k \tau_{\rm end}) ) }{2 k^2 \tau_{\rm end}^2 },\quad \beta_k = - \frac{i }{2 k^2 \tau^2_{\rm end}}~,
\end{align}
which satisfies the normalization condition $|\alpha_k|^2 - |\beta_k|^2 = 1$. 
%As first pointed out in ~\cite{Ford:1986sy}, the particle production given by the above process is finite but leads to a logarithmic divergent energy density. However, we will show that the logarithmic divergence is finite in reality and will never dominate the vacuum energy.
%This $\alpha_k$ and $\beta_k$ satisfies the normalization condition
%\begin{align}
%|\alpha_k|^2 - |\beta_k|^2 = 1~.
%\end{align}

%We can do an estimation of the energy density of the particles produced at the end of inflation to determine the range of validity of our result. Since for the mode function, we have used the subhorizon approximation. In the plot below, it means that our $k_{\rm min}$ is given by
%\begin{align}
%k_{\rm min} = (-\tau_{\rm end})^{-1}~.
%\end{align}

We have to ensure that the produced particle does not have much backreaction on our background described by Eq.~\eqref{equation622} and \eqref{equation623}. So we can estimate the minimum of $\Delta t$ by requiring that the energy density of the produced particles is comparable to the inflationary vacuum energy as follows:
%Since the inflationary vacuum energy density always remains the same, we only need to compare the energy density of the particle produced and the vacuum energy at the end of inflation. 
\begin{align}\nonumber
\rho_X & = \frac{1}{a^4(\tau_{\rm end})} \int 2 \pi k^3 d k \frac{1}{4 k^4 \tau_{\rm end}^4}  = \frac{2\pi}{a^4(\tau_{\rm end})} \log\bigg(\frac{k_{\rm max}}{k_{\rm min}}\bigg) \frac{1}{4 \tau_{\rm end}^4} \lesssim 3 M_{\rm pl}^2 H^2~.
\end{align} 
%Since $M_{\rm pl}^2/H^2$ is huge.
%\begin{align}
%M_{\rm pl} \sim 10^{19} {\rm GeV}, \quad H\sim 10^{16} {\rm GeV} \quad \rightarrow  \quad  M_{\rm pl}^2/H^2 \sim 10^6~.
%\end{align}
%We expect a very large $k_{\rm max}\sim e^{10^6}k_{\rm min}$:
%%\begin{align}
%%\frac{k_{\rm max}}{k_{\rm min}} \sim e^{10^6} ~.
%%\end{align}
%in order to introduce significant backreaction, which is beyond the Planck scale. So we conclude here that the backreaction  from the logarithmic divergence is negligible, and sudden transition from the inflation phase to radiation phase won't lead to any catastrophes. 
That is, as long as $\Delta t \gg m_X^{-1} e^{(\frac{H}{M_{\rm pl}})^2}$, backreaction is negligible. Now we estimate the particle number produced in this comoving $k$ ranging from $k_{\rm min}$ to $k_{\rm max}$ as
\begin{align}\nonumber
n_X & =\frac{1}{a(\tau_{\rm end})^3} \int_{k_{\rm min}}^{k_{\rm max}} d k 2\pi k^2 |\beta_{k}|^2   =\frac{1}{a(\tau_{\rm end})^3} \int_{k_{\rm min}}^{k_{\rm max}} 2 \pi k^2 d k \frac{1}{4 k^4 \tau_{\rm end}^4} \\
&=\frac{1}{a(\tau_{\rm end})^3} \frac{\pi}{2} \bigg(\frac{1}{k_{\rm min} }- \frac{1}{k_{\rm max}} \bigg) \frac{1}{\tau_{\rm end}^4}~.
\end{align}
%This is the number in a comoving volume, in order to obtain the number density, we can
%\begin{align}
%n_{X}=\frac{1}{a(\tau_{\rm end})^3} \frac{\pi}{2} \bigg(\frac{1}{k_{\rm min} }- \frac{1}{k_{\rm max}} \bigg) \frac{1}{\tau_{\rm end}^4} ~.
%\end{align}
Assuming $k_{\rm max}\gg k_{\rm min}$, $n_{X} \sim  \frac{H}{m_X} H^3 \sim H^3$ for $m_X \sim H$. No matter what value of $\mu$ we take, the de Sitter particle production is much less than the particle production from a sudden inflation-radiation transition era. The corresponding DM relic abundance can be estimated similarly:
%\begin{align}\nonumber
%\Omega_X h^2 & = 4.31\times 10^{-5} \frac{T_e}{T_0} \frac{\rho_X(t_e)}{\rho_I(t_e)}   = 4.31\times 10^{-5} \frac{T_e}{T_0} \frac{\frac{\pi }{2} H^3 m_X }{3M_{\rm pl}^2 H^2}  = 0.12~.
%\end{align}

\begin{align}\nonumber
\Omega_X h^2 & \simeq\frac{8\pi}{3}\Omega_{\text{R}}h^2 \frac{m_X n_X}{M_{\rm pl}^2 H^2}\bigg( \frac{T_{\rm RH}}{T_0} \bigg) \simeq 4.31\times 10^{-5} \frac{T_{\rm RH}}{T_0} \frac{\pi  H m_X }{6 M_{\rm pl}^2}  = 0.12~.
\end{align}
%We have $T_e=\sqrt{H M_{\rm pl}}$, $T_0\simeq 10^{-3}\, {\rm eV}$. Then we have, for $m_X\sim H = 10^8 {\rm GeV}$.
For instantaneous reheating case, we need $m_X\sim H \sim 10^8$~GeV to produce enough DM, which is much smaller compared to the ones in smooth transition scenarios as expected.

\section{Dark Matter on the Cosmological Collider} \label{collider}

Let us consider an effective field theory where the dark matter sector couples to the primordial curvature perturbation $\zeta$ in the following way \cite{An:2018tcq}:
\begin{align}\label{twoactions}
	S_{1} = c_1\int d^4 x a^3 \zeta' X^2, \quad S_{2} = c_2\int d^4x a^2 \zeta'\zeta' X^2~.
\end{align}
These two operators can be regarded as coming from the effective field theory of inflation, or two terms coming from the full Lagrangian of the original quasi-single field inflation with a constant turning trajectory \cite{Chen:2009zp}. Since if we consider other types of interactions, the procedure of obtaining the characteristic signals will be similar. Thus, we focus on these two terms for simplicity. 

Due to the above interactions, the DM leaves its imprints in the primordial non-Gaussianity. Here let us assume that $m_X\gg H$. The dominant contribution to the non-Gaussianity is from integrating out the dark matter field \cite{Chen:2012ge,Pi:2012gf,Tong:2017iat,Iyer:2017qzw,Wu:2018lmx}. The resulting non-Gaussianity is of the shape of general single field inflation \cite{Chen:2006nt} and of equilateral shape. The magnitude of non-Gaussianity is power-law suppressed in the large-mass limit \cite{Gong:2013sma}. 

\begin{figure}[htbp] 
	\centering 
	\includegraphics[width=5cm]{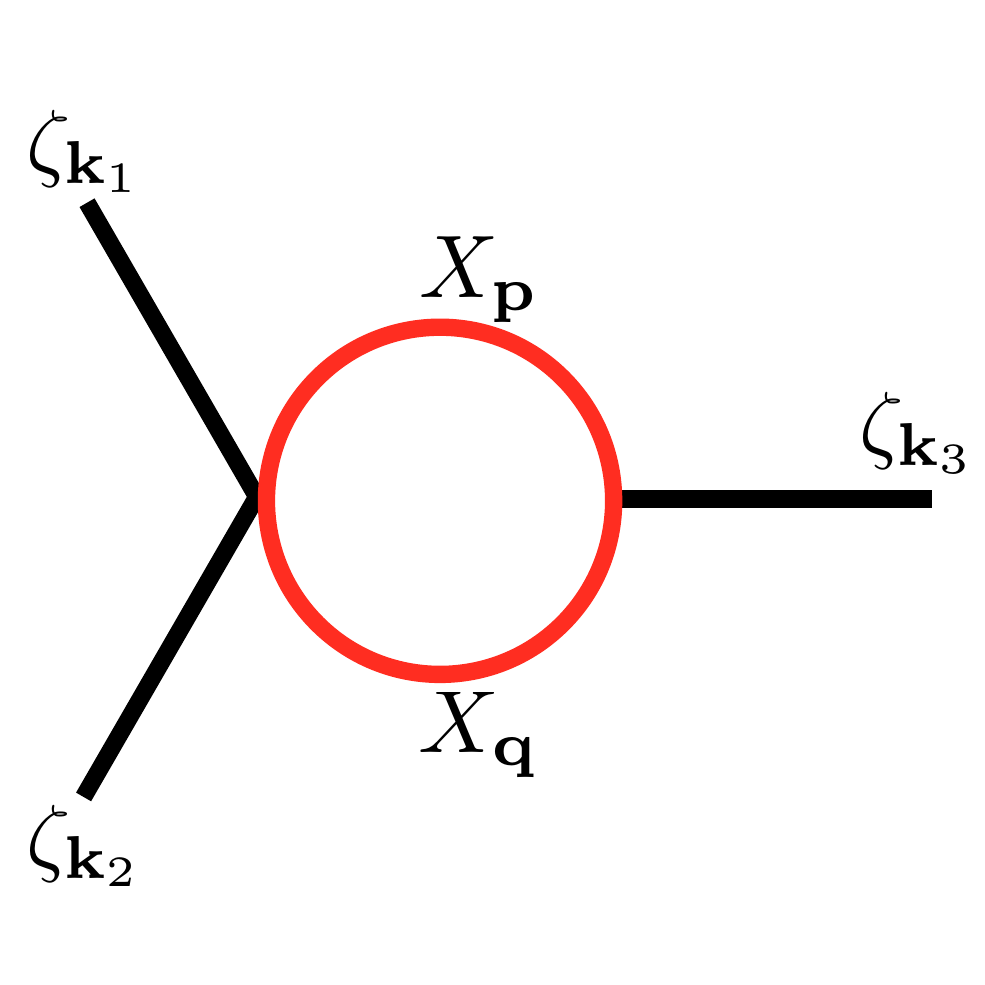} \caption{The Feynman diagram that contributes to the cosmological collider signal in the squeezed limit $k_3\ll k_1\sim k_2$.}   \label{loopngf}
\end{figure}

There is another type of signal, which, in magnitude, is smaller than the equilateral shape non-Gaussianity from integrating out heavy fields. Usually they are subject to Boltzmann suppression $e^{-\pi \mu}$ in the large mass limit. This signal, however, is more informative and can tell us directly the mass of the dark matter field in terms of Hubble rate during inflation. This is the cosmological collider signal in the squeezed limit non-Gaussianity from which we can also measure the spin of the dark matter particle. This mechanism is known as the cosmological collider \cite{Arkani-Hamed:2015bza} and is closely related to quasi-single field inflation \cite{Chen:2009we,Chen:2009zp,Baumann:2011nk} and the primordial quantum standard clocks \cite{Chen:2015lza,Chen:2016cbe,Chen:2016qce}. 

Given the form of interactions, the $\langle \zeta_{\mathbf k_1}\zeta_{\mathbf k_2}\zeta_{\mathbf k_3} \rangle'$ (the prime indicates that we ignore the momentum-conservation factor $(2\pi)^3\delta^3(\mathbf k_1+\mathbf k_2+\mathbf k_3)$) is contributed by the loop diagrams. The dark matter loop can be attached to any of the $\mathbf k_1$, $\mathbf k_2$ and $\mathbf k_3$ legs. In the squeezed limit $k_3\ll k_{1}\sim k_2$, the leading contribution to the cosmological collider signal comes from the Feynman diagram depicted in FIG.~\ref{loopngf}. The technique of computing this type of loop diagram is initially proposed in \cite{Arkani-Hamed:2015bza}. The application to the Standard model mass spectrum can be found in \cite{Chen:2016nrs,Chen:2016uwp,Chen:2016hrz,Wu:2017lnh}, and it had also been applied to complex scalar fields in \cite{Chua:2018dqh}. The crucial point is that although in general, loop diagrams may suffer from UV or IR divergence \cite{Weinberg:2010wq}, and usually we need to introduce local counter terms to cancel the UV divergence, the clock-signal part of the contribution does not suffer from UV or IR divergence. The reason is that the contribution is coming from the non-local process. In this case, we only have to use a double Fourier transformation to deal with the momentum integral. We present the details of the computation in Appendix \ref{loopdetails}. 
\begin{figure}[htbp] 
	\centering 
	\includegraphics[width=9cm]{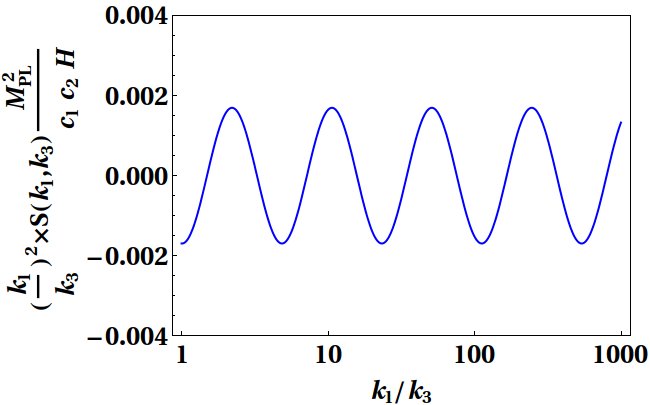} \caption{The cosmological collider signal $(k_1/k_3)^2\times S(k_1,k_3)\times(\frac{M_{PL}^2}{c_1c_2H^2})$ as a function of the ratio $k_1/k_3$ with $\mu=2$. }   \label{loopng1}
\end{figure}
It gives the standard cosmological collider signal in the squeezed limit as
\begin{align}\label{clocksignals}
	\langle \zeta_{\mathbf k_1}\zeta_{\mathbf k_2}\zeta_{\mathbf k_3} \rangle' = c_1 c_2 {\rm Re} \bigg[ g(\mu) \frac{2^{-1-8i\mu}H^5}{k_1 k_2(k_1+k_2)^4 M_{\rm pl}^6 \epsilon^3 } \bigg(\frac{k_1+k_2}{k_3}\bigg)^{2i\mu} \bigg]~,
\end{align}
where the factor $g(\mu)$ is 
\begin{align}
	g(\mu) = \frac{\Gamma(2-2i\mu)\Gamma(4-4i\mu)\Gamma(-2i\mu)^4}{\Gamma(1/2-i\mu)^2 \Gamma(1/2+i\mu)^2} \sinh^2(\pi\mu)~.
\end{align}
We can express the bispectrum in terms of the dimensionless shape function $S(k_{1},k_{2},k_{3})$ \cite{Chen:2018xck} as
\begin{align} \label{shapefunction}
\langle \zeta_{\mathbf k_1} \zeta_{\mathbf k_2}  \zeta_{\mathbf k_3}  \rangle'   \equiv (2 \pi)^4 S(k_{1},k_{2},k_{3}) \frac{1}{(k_{1}k_{2}k_{3})^2} P^{(0)2}_{\zeta}~,
\end{align}
where $P^{(0)}_\zeta = \frac{1}{8\pi^2 {M_{\rm pl}^2}} \frac{H^2}{\epsilon}$ is the power spectrum of the curvature perturbation without the correction caused by massive fields. 
We are particularly interested in the squeezed limit where $k_1\sim k_2\gg k_3$. In this limit, the shape function is
\begin{align}
S(k_1,k_3) = c_1 c_2 {\rm Re} \bigg[ g(\mu) \frac{2^{3-8i\mu} H}{M_{\rm pl}^2\epsilon} \bigg(\frac{2k_1}{k_3}\bigg)^{2 i \mu-2} \bigg] ~,
\end{align}
and we plot $k_1/k_3\times S(k_1,k_3)$ in FIG.~\ref{loopng1}.

The estimator of non-Gaussianity can be defined as \cite{Chen:2016hrz}
\begin{align}
	f_{\rm NL} \simeq 8c_1c_2\bigg|g(\mu)\frac{H}{M_{\rm pl}^2 \epsilon} \bigg|.
\end{align} 

The modest lower bound of $f_{\rm NL}$ can be estimated for an effective field theory with Planck scale cutoff. For such a theory, $c_1\sim H$ and $c_2\sim 1$ (note that a $1/(\sqrt{\epsilon} M_p)$ factor emerges when we canonically normalize $\zeta$). This will lead to vanishingly small non-Gaussianities. For the toy universe of smooth transition, the typical value for $f_{\rm NL}$ is $10^{-24}$ for $r=0.05$, $H\sim 10^{13}$~GeV and $\mu\simeq 4.9$. For $r=0.001$, $H\sim 10^{12}$~GeV and $\mu\simeq 4.2$, $f_{\rm NL} \sim 10^{-22}$. For the instantaneous case, the typical value for $f_{NL}$ is $10^{-10}$ for the parameters we considered in the last section.  

For the effective field theories which are not Planck mass suppressed, $c_1$ can be larger than $H$, $c_2$ can be larger than $1$. But too large couplings with the inflaton may change the relic abundance. To estimate the upper bound, and thus cause inconsistencies for the dark matter production. 

Note that the two operators in \eqref{twoactions} actually originates from the following two actions
\begin{align}\label{action111}
\tilde S_{1} = \frac{\tilde c_1}{\Lambda} \int d^4 x a^3  \phi' X^2, \quad \tilde S_{2} = \frac{\tilde c_2}{\Lambda^2} \int d^4x a^2  \phi' \phi' X^2~.
\end{align}
where $c_1$ and $\tilde c_1$, $c_2$ and $\tilde c_2$ are related via
\begin{align}
	\tilde c_1 = \frac{c_1}{\sqrt{\epsilon} M_{\rm pl}}, \quad \tilde{c}_2  = \frac{\Lambda^2 c_2}{\epsilon M_{\rm pl}^2}~.
\end{align}

Now we want to consider $\tilde c_1$ and $\tilde c_2$ as order one constants. The bound on the scale of the EFT should be given by the constraint that there are no overproduction of the dark matter relic abundance. Note that the decay chanel from a single inflaton to dark matter is kinematically forbidden since the inflaton is lighter than the dark matter. Thus, to check the relic abundance, the leading contribution comes from the dark matter produced by collisions of the thermalized inflaton particles.

The explicit form of the constraint depends on the details of reheating. If the inflaton is thermalized rather efficiently during reheating, to get the correct relic abundance for dark matter, we need
\begin{align}
	\Lambda \gtrsim 10^{1.5} \bigg(\frac{M_{\rm pl}}{m_{\phi}} \frac{m_X}{1{\rm GeV}}\bigg)^{\frac{1}{4}} T_{\rm RH}~.
\end{align}
For the typical parameter space, we have $\Lambda \geq 10^5 T_{\rm RH}$. Depending on the reheating temperature, $\Lambda$ can take different values. 

On the other hand, if the inflaton has no chance to be thermalized, the above bound does not apply. For example, one may consider brane inflation for the sudden transition \cite{Dvali:1998pa,HenryTye:2006uv}, where the inflaton just disappears
at the moment of reheating. In addition, for smooth transition, one may consider quintessential inflation, where reheating can also be realized by gravitational particle production, without assuming the coupling between the inflaton and the standard model sector.
In principle, dark matter can be created from the inflaton during kination as well, but during kination the energy density of the inflaton gets redshifted rapidly $\propto a^{-6}$, and hence this production would not cause a problem for our estimations.
One can also consider a scenario where the inflaton decays to a heavy field, while dark matter is created gravitationally at the end of inflation, and the former heavy field may later decay to radiation to reheat the Universe. In this case the reheating temperature
is determined by the decay rate of the heavy field, and then the reheating temperature can be chosen to be sufficiently low so that dark matter cannot get produced at the reheating. In general, either way, we need to discuss gravitational particle production
by a smooth transition from inflation to kination or an early matter domination, and applying the Stokes line method to such situations would also be possible, though we used a smooth transition from inflation to Minkowski spacetime for simplicity. In gravitational
reheating one also needs to worry about an overproduction of gravitons, but this issue can be circumvented by spinodal instability of the Higgs or by introducing a sufficient number of degrees of freedom for radiation gravitationally produced \cite{Nakama:2018gll}. One can also use a scenario of \cite{Hashiba:2018tbu}.
We estimated the relic abundance assuming smooth transition from inflation to radiation domination, but an extension of that analysis to cases with a kination phase inserted would be straightforward.

Another lower bound on $\Lambda$ (and thus upper limit on $f_{\rm NL}$) can be obtained from the consideration that \eqref{action111} does not change the dark matter mass too much (otherwise we lose the prediction power for the mass of the dark matter particle on the cosmological collider). For instance, if $\xi\sim 1$ and $R\sim H^2$ this means if $\Lambda > 10^4H$, then this additional term is negligible. Otherwise, the mass of dark matter on the cosmological collider will significantly differ from the probes that we expect to use to observe dark matter now. This limit will generically give $f_{\rm NL}<10^{-6}$. Thus, we expect that if the cosmological collider signal is observed for such dark matter production scenarios, it is very likely that the observed mass of the dark matter is strongly corrected by the inflaton coupling. However, this conclusion depends on the coupling details between the inflaton and the dark matter. Since we haven't exhausted all possible couplings between the inflaton and dark matter, rather only studied the simplest ones, it remains interesting to see if there can be dark matter couplings with the inflaton which can naturally take large values.

The dark matter field may also leave some imprints on the power spectrum through the relation which relates $\langle \zeta \zeta X^2 \rangle $ and $\langle \zeta \zeta \rangle$ when the comoving momentum of $\zeta$ is soft \cite{Chung:2013sla}. However, these relations may not show characteristic features of dark matter, but rather are universal for all matter components \cite{Pajer:2013ana}. For example, one can expand the correlation function in the series of the ratio between the soft leg and the hard leg. The leading order gives the Maldacena consistency relation which is fixed by dilatation symmetry \cite{Maldacena:2002vr}. The next-to-leading order is fixed by special conformal symmetry \cite{Creminelli:2004yq}. It is thus unclear how to extract dark mater properties from these correlators.

Finally, we briefly comment possible isocurvature fluctuations from dark matter. A field like $X$ that is not directly coupled to the inflaton will introduce isocurvature perturbation. From ref.~\cite{Chung:2004nh}, we can estimate the size of $P_{\delta X}$ as
\begin{equation}
P_{\delta X} \sim \frac{k^3 m_X^4}{2 \pi^2 \rho_X^2}\int d^3 r e^{i \vec{k} \cdot \vec{r}} \left\langle X(\vec{x}) X(\vec{y}) \right\rangle^2~.
\end{equation}
Since $X$ is a heavy field ($\mu >$  0), $X(k)\propto (\frac{k}{a H})^{i \mu}$, which has no significant $k$ dependence. At large scale, $k\to 0$, $P_{\delta X}$ will be suppressed by $k^3$. In other words, the power of isocurvature mode is diluted by inflation with $\gtrsim 40$ e-folds. Therefore we do not expect any large scale isocurvature signal to be observed.

\section{Conclusion}\label{summary}
We considered three cosmological scenarios and calculated the dark matter relic abundance produced gravitationally. Our conclusions are as follows: For exact de Sitter space, the number density is proportional to $\frac{H^3 \mu^3 2\pi}{3e^{2\pi\mu}-1}$, where $\mu\equiv\sqrt{m_X^2/H^2-9/4}$. For a universe where inflation is connected to a Minkowski universe, we used the Stokes line method to calculate the dark matter relic abundance. The resulting number density is exponentially suppressed. We fit the results numerically by Eq.~\eqref{empirical}. For a sudden transition where inflation is immediately connected to a radiation-dominated universe, the produced particle number density is proportional to $H^3$.

The cosmological collider signal of the dark matter particle is discussed. We note that with the simplest couplings between the inflaton and dark matter, for the inflaton coupling to satisfy two conditions: (1) does not overproduce dark matter and (2) does not correct the dark matter mass significantly, the non-Gaussianity produced is too small to observe in future observations. It is interesting to see if there are mechanisms to boost the non-Gaussianity of the scenario to observational range to test the scenarios of gravitationally produced superheavy dark matter.

\section*{Acknowledgments}  
We thank Xingang Chen, Yohei Ema and Zhong-Zhi Xianyu for useful discussions on general mechanisms for gravitational particle production. This work is supported in part by ECS Grant 26300316 and GRF Grant 16301917 and 16304418 from the Research Grants Council of Hong Kong.  

\appendix

\section{Particle Production from the Divergent Asymptotic Series Method}
\label{sec:app-div}

In this appendix, we review the divergence series method and use it to study the particle production for general FRW backgrounds. We will first summarize the result in \ref{sec:app-div-summary}. The summarized result will then be derived in \ref{sec:app-div-derivation}. In \ref{sec:app-div-example}, as a simple example, we use this method to recover the known results for particle production in de Sitter space.

\subsection{Summary of the Result}
\label{sec:app-div-summary}

To study the particle production problem, we start from the equation of motion of the massive field, which can be rewritten as
\begin{equation}
\ddot{f}_{k}+\omega_{k}^{2}(t)f_{k}=0,~
\omega_{k}(t)=\sqrt{\frac{k^{2}}{a^{2}}+m^{2}-\frac{9}{4}H^{2}-\frac{3}{2}\dot{H}}\ .
\label{eq:EOM}
\end{equation}
This equation, together with the instant Minkowski initial condition, defines the problem of particle production on FRW backgrounds.

As inspired by scattering problems in quantum mechanics, the sign of $\omega^2_k(t)$ determines whether the mode function is oscillating or decaying in a ``potential barrier'', and the moments when $\omega_k(t)=0$ are turning points. For large enough mass $m$, the turning points can be complex. The phase integral along a particular complex line crossing the real axis is connected to an exponentially small component in the mode function \cite{Berry1989}. 

We denote $t_{c}$ as the complex time located at lower half-plane which makes $\omega_{k}(t_{c})=0$, together with a phase accumulated from $t_c$
\begin{equation}
\Theta_{k}(t)=-i\int_{t_{c}}^{t}\omega_{k}(t')dt'\ .\label{eq:phase}
\end{equation}
The contour to carry out the integral will be specified in Eq.~\eqref{eq:refstokes}.

We define the Dingle's singulant variable
\begin{equation}
F_{k}(t)=2\Theta_{k}(t)\ .\label{eq:singulant}
\end{equation}
With these definitions, we write down the form of the approximate
solution of Eq.~(\ref{eq:EOM}) \cite{Berry1989}:
\begin{align}
f_{k}(t) &\approx \mathrm{exp}[-\Theta_{k}(t_{i})]\frac{\mathrm{exp}[\Theta_{k}(t)]+iS_{k}(t)\mathrm{exp}[-\Theta_{k}(t)]}{\sqrt{2\omega_{k}}}\nonumber\\
&=\frac{1}{\sqrt{2\omega_{k}}} \Big \{ \mathrm{exp} \Big(-i\int_{t_{i}}^{t}\omega_{k}dt' \Big)-iS_{k}(t)\mathrm{exp}[-F_{k}(t_{i})]\mathrm{exp}\Big(i\int_{t_{i}}^{t}\omega_{k}dt' \Big) \Big \} \ .
\label{eq:approximation}
\end{align}
where $S_{k}$ is called the Stokes multiplier function 
\begin{align}
S_{k}(t)=\frac{1}{2}\Big[1+\mathrm{Erf}\Big(\frac{-\mathrm{Im}F_{k}(t)}{\sqrt{2|\mathrm{Re}F_{k}(t)|}}\Big)\Big ]~. \label{S_k_expression}
\end{align}
The constant factor $\mathrm{exp}[-\Theta_{k}(t_{i})]$ in the first line of Eq.~(\ref{eq:approximation})
is for matching the adiabatic vacuum $\frac{1}{\sqrt{2\omega_{k}}}e^{-i\int_{t_{i}}^{t}\omega_{k}dt'}$
when $S_{k}\to0$, and the amplitude of the negative-frequency part is then suppressed by the exponential
\begin{align}
\mathrm{Re}[F_{k}(t_i)] &=\mathrm{Re}\Big[-2i\int_{t_{c}}^{t_{m}}\omega_{k}dt-2i\int_{t_{m}}^{t_{i}}\omega_{k}dt \Big] =-i\int_{t_{c}}^{t_{c}^{*}}\omega_{k}dt~. \label{eq:refstokes}
\end{align}
The moment when $\mathrm{Im}F_k(t)=0$ corresponds to the emergence of the negative-frequency part of the mode function, and the set of complex $t$ satisfying this condition forms the Stokes line \cite{Berry1989}. In the above equation, $t_m$ is where the Stokes line intersects the real time axis. In the integration, we first integrate from $t_c$ to $t_m$ along the Stoke line, then integrate from $t_m$ to t along the real axis.

From Eq.~(\ref{eq:approximation}), the approximated Bogoliubov coefficients can be extracted as
\begin{align}
\alpha_{k}(t)\approx1,~\beta_{k}(t)\approx-iS_{k}(t)\mathrm{exp}\Big( i\int_{t_{c}}^{t_{c}^{*}}\omega_{k}dt' \Big)~,
\label{eq:Bogoliu}
\end{align}
which agrees with the results shown in \cite{Dumlu2010,Dabrowski2014,Dabrowski2016}. 

\subsection{The Divergent Asymptotic Series Method}
\label{sec:app-div-derivation}

As suggested by Dingle \cite{Dingle1973} and Berry \cite{Berry1989}, to get the particle production in \eqref{eq:approximation}, we assume that the dominant part of
Eq.~(\ref{eq:EOM}) has an asymptotic series solution $y$ in terms of a large parameter $m$ (or using $\mu=\sqrt{m^2-9H^2/4}$ in the cases with constant $H$).
\begin{align}
y_{k}(t) & =\frac{\mathrm{exp}[\Theta_{k}(t)]}{\sqrt{\omega_{k}}}\sum_{j=0}^{\infty}b_{k}^{(j)} =\frac{\mathrm{exp}[\Theta_{k}(t)] }{\sqrt{\omega_{k}}}B_{k}\ ,\label{eq:series1}
\end{align}
where $b_k^{(j)}\propto m^{-j}$ is the $j$-th order correction. Substitute the series in Eq.~(\ref{eq:EOM}), yielding
\begin{equation}
\ddot{B}_{k}-\Big(2i\omega_{k}+\frac{\dot{\omega}_{k}}{\omega_{k}}\Big)\dot{B}_{k}+\frac{3\dot{\omega}_{k}^{2}-2\omega_{k}\ddot{\omega}_{k}}{4\omega_{k}^{2}}B_{k}=0\ .\label{eq:Bequ}
\end{equation}
It is convenient to introduce a variable $W_{k}(t)=\omega_{k}^{2}(t)$
to investigate the analytic property of the differential equation.
With this substitution, Eq.~(\ref{eq:Bequ}) becomes
\begin{equation}
\ddot{B}_{k}-\Big(2iW_{k}^{\frac{1}{2}}+\frac{\dot{W}_{k}}{2W_{k}}\Big)\dot{B}_{k}+\frac{5\dot{W}_{k}^{2}-4W_{k}\ddot{W}_{k}}{16W_{k}^{2}}B_{k}=0\ .
\label{eq:substituteX}
\end{equation}
By rearranging the equation and integration by part, Eq.~(\ref{eq:substituteX})
generates
\begin{align}
iB_{k} & =\int^{t}\frac{\ddot{B}_{k}}{2W_{k}^{\frac{1}{2}}}-\frac{\dot{W}_{k}\dot{B}_{k}}{4W_{k}^{\frac{3}{2}}}dt'+\int^{t}\frac{5\dot{W}_{k}^{2}-4W_{k}\ddot{W}_{k}}{32W_{k}^{\frac{5}{2}}}B_{k}dt'\nonumber \\
& =\frac{\dot{B}_{k}}{2W_{k}^{\frac{1}{2}}}+\int^{t}\frac{5\dot{W}_{k}^{2}-4W_{k}\ddot{W}_{k}}{32W_{k}^{\frac{5}{2}}}B_{k}dt'\ .\label{eq:integralform}
\end{align}
Using the fact that $W_k $ corresponds to $\mathcal{O}(m^2)$ and matching the orders of $m$ in the series (\ref{eq:series1}), we obtain a recurrence relation for $b_{k}^{(j)}$:
\begin{equation}
ib_{k}^{(j+1)}=\frac{\dot{b}_{k}^{(j)}}{2W_{k}^{\frac{1}{2}}}+\int^{t}\frac{5\dot{W}_{k}^{2}-4W_{k}\ddot{W}_{k}}{32W_{k}^{\frac{5}{2}}}b_{k}^{(j)}dt'\ .\label{eq:recurrence}
\end{equation}
Since the series is expanded around $t_{c}$, it is reasonable to
investigate the recurrence relation around this point. Assuming that
$W_{k}$ is analytical around $t_{c}$, we can expand in terms of $(t-t_c)^n$:
\begin{equation}
W_{k}(t)=\dot{W}_{k}(t_{c})(t-t_{c})+\frac{\ddot{W}_{k}(t_{c})}{2}(t-t_{c})^{2}+\mathcal{O}(|t-t_{c}|^{3})\ .\label{eq:expansion}
\end{equation}
By defining a variable
\begin{align}
q_{k}(t) & =\int^{t}W_{k}^{-\frac{3}{2}}\ddot{W}_{k}dt' 
 \approx-2\ddot{W}_{k}(t_{c})\dot{W}_{k}(t_{c})^{-\frac{3}{2}}(t-t_{c})^{-\frac{1}{2}} 
 =C(t-t_{c})^{-\frac{1}{2}}\ ,\label{eq:q}
\end{align}
Eq.~(\ref{eq:recurrence}) reduces to a polynomial differential
equation
\begin{equation}
-4i\dot{W}_{k}(t_{c})^{\frac{1}{2}}C^{3}b_{k}^{(j+1)}\approx q_{k}^{4}\frac{db_{k}^{(j)}}{dq_{k}}+\int_{0}^{q_{k}}\frac{5q_{k}^{2}}{4}b_{k}^{(j)}dq_{k}'\ ,\label{eq:polynomialq-1}
\end{equation}
where we keep only the dominant term in the integrand. This recurrence relation can be solved by applying an ansatz $b_{k}^{(j)}=\tilde{b}_{k}^{(j)}q_{k}^{3j}$
\begin{equation}
-4i\dot{W}_{k}(t_{c})^{\frac{1}{2}}C^{3}\tilde{b}_{k}^{(j+1)}=\Big(3j+\frac{5}{12j+12}\Big)\tilde{b}_{k}^{(j)}\ .\label{eq:differenceequ}
\end{equation}
Let $\tilde{b}_{k}^{(0)}=1$, the approximation for
$b_{k}^{(j)}$ near $t_{c}$ is then given by
\begin{equation}
b_{k}^{(j)}\approx\Big[\frac{3q_{k}^{3}}{-4i\dot{W}_{k}(t_{c})^{\frac{1}{2}}C^{3}}\Big]^{j}\frac{\Gamma(j+\frac{5}{6})\Gamma(j+\frac{1}{6})}{2\pi\Gamma(j+1)}\ .\label{eq:bj}
\end{equation}
This solution is quite complicated, but it is enough to know the divergent behavior when $j\to\infty$. We rewrite Eq.~(\ref{eq:bj}) in terms of the singulant variable in the large-$j$ limit as
\begin{align}
b_{k}^{(j)}(t) & \approx\frac{(j-1)!}{2\pi}\frac{1}{(-2i\int_{t_{c}}^{t}W_{k}^{\frac{1}{2}}dt')^{j}} =\frac{(j-1)!}{2\pi F_{k}(t)^{j}}\ .\label{eq:largeterm}
\end{align}
One can easily check that $b_k^{(j)} \propto \mathcal{O} (m^{-j})$ which agrees with the ansatz of asymptotic series \eqref{eq:series1}, and this equation indicates that the series $B_{k}$ is divergent in factorial form. We will show that the factorial divergence is crucial since it is related to the Borel summation which generates a subdominant term in the solution.

Now we apply Berry's theory \cite{Berry1989} to derive
the Stokes multiplier function $S_{k}(t)$ shown in  Eq.~\eqref{S_k_expression} which describes the moment when massive particles emerge. To handle a divergent asymptotic series such as Eq.~(\ref{eq:series1}),  the standard method \cite{Higham2015} is to truncate the series into a finite sum and a divergent tail as

\begin{equation}
y_{k}(t)\approx\frac{\mathrm{exp}[\Theta_{k}(t)]}{\sqrt{\omega_{k}}}\sum_{j=0}^{n-1}b_{k}^{(j)}-\frac{i\mathrm{exp}[-\Theta_{k}(t)]}{\sqrt{\omega_{k}}}\Big[\frac{i}{2\pi}e^{F_{k}(t)}\sum_{j=n}^{\infty}\frac{(j-1)!}{F_{k}(t)^{j}}\Big]\ ,\label{eq:truncate}
\end{equation}
assuming $n$ is large enough so that Eq.~(\ref{eq:largeterm}) is applicable. By comparing with Eq.~(\ref{eq:approximation}), the terms inside the square bracket is the Stokes multiplier, and we then apply the Borel summation to make the series sum meaningful. The first step is to convert the factorial into the integral of Gamma function, and we denote the multiplier as

\begin{align}
S_{k}(t) & =\frac{i}{2\pi}e^{F_{k}(t)}\int_{0}^{+\infty}ds\frac{e^{-s}}{s}\sum_{j=n}^{\infty}\Big(\frac{s}{F_{k}(t)}\Big)^{j} =\frac{i}{2\pi}\int_{0}^{+\infty}ds\frac{e^{F_{k}(1-\frac{s}{F_{k}})}}{s}\Big(\frac{s}{F_{k}}\Big)^{n}\frac{1}{1-\frac{s}{F_{k}}}\nonumber \\
& =-\frac{i}{2\pi}\int_{-1}^{+e^{-\mathrm{Arg(F_{k})}}\infty}dz\frac{(1+z)^{n-1}e^{-F_{k}z}}{z} = -\frac{i}{2\pi}\int_{-1}^{+\infty} dz\frac{e^{(n-1)\log(1+z)-F_{k}z}}{z},\label{eq:resummation1}
\end{align}
where the variable changes to $z=\frac{s}{F_{k}(t)}-1$ in the last
line, and the contour is deformed that the upper limit is $+\infty$.
%\begin{figure}
%\includegraphics[scale=0.4]{U_Contour}
%\caption{The contour $U$ for the integration of $S_k$.} \label{fig:U_Contour}
%\end{figure}
The denominator $z$ indicates
that the magnitude of the integrand dominates at $z\approx0$, and
we can choose $n\approx |F_k(t)|+1$ such that the phase is stationary at $z\approx0$,
implying that the integral can be evaluated with saddle-point approximation. Before proceeding the calculation of $S_k(t)$, we first justify two conditions required by the validity of the saddle-point approximation: $|F_k(t)|$ has to be sufficiently close to $|\mathrm{Re}F_k(t)|$ since adjusting $n$ can only cancel the real part of the exponent, and $|\mathrm{Re}F_k(t)|$ has to be large enough such that Eq.~\eqref{eq:largeterm} is applicable. The first condition is valid when the evolution is close to the Stokes line when $\mathrm{Im}F_k(t)=0$, implying that the saddle point approximation is valid around the particle production, and it is sufficient to explain the emergence of the sub-dominant term in the mode function. To justify the second condition, we use the fact that  $\mathrm{Re}F_k(t)$ is a constant on real axis and proportional to the large parameter $m$. For the cases we considered in this paper, $\mathrm{Re}F_k(t)$ is sufficiently large as long as $m \gtrsim H$.

With the notations $F_{k}^{R}={\rm Re}F_{k}$ and $F_{k}^{I}={\rm Im}F_{k}(t)$, we expand the integrand in Eq.~\eqref{eq:resummation1} around $z=0$ and approximate the integral from $-\infty$ to $+\infty$:
\begin{align}
S_{k}(t) & \approx-\frac{i}{2\pi}\int_{-\infty}^{+\infty}dz\frac{e^{(n-1-F_{k})z-\frac{n-1}{2}z^{2}}}{z}\nonumber \\
& \approx -\frac{1}{2\pi}\int_{-\infty}^{+\infty}dz\frac{\mathrm{sin}(F_{k}^{I}z)+i\mathrm{cos}(F_{k}^{I}z)}{z}e^{-\frac{n-1}{2}z^{2}}\Big[1+(n-1-F_{k}^{R})z\Big]\nonumber \\
& =-\frac{i}{2\pi}\int_{-\infty}^{+\infty}dz\frac{\mathrm{cos}(F_{k}^{I}z)}{z}e^{-\frac{n-1}{2}z^{2}}-\frac{1}{2\pi}\int_{-\infty}^{+\infty}dz\frac{\mathrm{sin}(F_{k}^{I}z)}{z}e^{-\frac{n-1}{2}z^{2}} \nonumber \\
&-i\frac{(n-1-F_{k}^{R})}{2\pi}\int_{-\infty}^{+\infty}\mathrm{cos}(F_{k}^{I}z)e^{-\frac{n-1}{2}z^{2}}\nonumber \\
& = S_{-}-\frac{1}{2}\mathrm{Erf}\Big[\frac{F_{k}^{I}}{\sqrt{2(n-1)}}\Big]-i\frac{(n-1-F_{k}^{R})}{\sqrt{2\pi(n-1)}}e^{-(F_{k}^{I})^{2}/(2n-2)}\nonumber \\
& \approx S_{-}+\frac{1}{2}\mathrm{Erf}\Big[-\frac{F_{k}^{I}}{\sqrt{2 F_{k}^R}}\Big]\ ,\label{eq:multiplier}
\end{align}
where the constant
\begin{align}
S_{-}=-\frac{i}{2\pi}\int_{-\infty}^{+\infty}dz\frac{\mathrm{cos}(F_{k}^{I}z)}{z}e^{-\frac{n-1}{2}z^{2}}~,
\end{align}
and we apply the fact in the last
line that $|n-1-F_{k}^{R}|\lesssim \mathcal{O}(1)$ as long as $n$ is sufficiently large. Physically, the initial condition requires $S_k(t)=0$ when $t$ is small, which indicates that the correct $S_{-}$ should be $1/2$. In the original method~\cite{Berry1989} this is handled by taking the principal value of the integral. Here we show that one can also evaluate $S_{-}$ by deforming the contour near the origin to be an infinitesimally small semicircle below the singularity $z=0$. Since the integrand is odd, only the integral over the semicircular contour contributes to $S_{-}$~\cite{Wong1990}:
\begin{align}
S_{-}&= \lim_{\delta\to 0^+}\frac{1}{2\pi}\int_{\pi}^{2\pi} d\theta ~\mathrm{cos}(F_{k}^{I}\delta e^{i\theta})e^{-\frac{n-1}{2}\delta^2 e^{2i\theta}}=\frac{1}{2}~.
\end{align}
Thus $S_k(t)$ indeed satisfies the initial condition $S_k(t) \sim 0$ when $t$ is small and $F_k^I(t) \ll 0$. Moreover, Eq.~\eqref{eq:multiplier} also indicates that the particle productions of massive
fields are governed by the error function, and the moment of productions
is represented by the time $t_m$ which satisfies $F_{k}^{I}(t_m)=0$. 

\subsection{Example: Particle Production in de Sitter}
\label{sec:app-div-example}

We now show that Eq.~(\ref{eq:Bogoliu}) does provide us a reasonable estimation of the Bogoliubov coefficients in inflation scenario $a(t)=e^{Ht}$. For the massive field theory with $m>\frac{3}{2}H$, Eq.~(\ref{eq:EOM}) becomes
\begin{equation}
k^{2}e^{-2Ht}+\mu^{2}H^{2}=0\ ,\label{eq:inflationcond}
\end{equation}
where $\mu=\sqrt{\frac{m^{2}}{H^{2}}-\frac{9}{4}}$, and the complex solutions which are closest to real axis are given by
\begin{equation}
t_c=-\frac{\log\frac{\mu H}{k}}{H}- i\frac{\pi}{2H}\ .\label{eq:solutions}
\end{equation}
Given the exact phase integral
\begin{align}
\Theta_{k}(t) & =-i\int_{t_{c}}^{t}\omega_{k}dt' =-i\Big[H\mu(t-t_{c})-\frac{\omega_{k}(t)}{H}+\mu\log\Big(1+\frac{\omega_{k}(t)}{H\mu}\Big)\Big]\nonumber \\
& =\frac{\pi\mu}{2}-i\Big[H\mu\Big(t+\frac{\log\frac{\mu H}{k}}{H}\Big)-\frac{\omega_{k}(t)}{H}+\mu\log\Big(1+\frac{\omega_{k}(t)}{H\mu}\Big)\Big]\ ,\label{eq:phaseinf}
\end{align}
we can derive $\beta_k$ in the inflation scenario
\begin{align}
\beta_k\approx -i\mathrm{exp}\Big(i\int_{t_{c}}^{t_{c}^{*}}\omega_{k}dt'\Big) S_k(t) 
=-ie^{-\pi\mu}S_k(t)\ ,\label{eq:Boltzmann}
\end{align}
which agrees with the definition of de Sitter temperature, and
the Stokes multiplier is
\begin{equation}
S_{k}(t)=\frac{1}{2}\Big[1+\mathrm{Erf}\Big(\frac{2H\mu\Big(t+\frac{\log\frac{\mu H}{k}}{H}\Big)-\frac{2\omega_{k}(t)}{H}+2\mu\log\Big(1+\frac{\omega_{k}(t)}{H\mu}\Big)}{\sqrt{2\pi\mu}}\Big)\Big]\ .\label{eq:multiplier-inf}
\end{equation}
For the mode with smaller $k$, it leaves horizon earlier. This is indicated by larger value of $\log\frac{\mu H}{k}$ which
makes the first term in the error function of Eq.~(\ref{eq:multiplier-inf}) dominates. The numerical solution of the zero imaginary part of Eq.~(\ref{eq:phaseinf}) shows that the typical production time $t_m(k)$ of each mode satisfies the relation
\begin{equation}
\frac{k}{a(t_m(k))\mu H}\approx0.66\ .\label{eq:ptinfexact}
\end{equation}

\section{Details of the Computation of the Loop Diagram}\label{loopdetails}
In this section, we present the details of the derivation of the cosmological collider signal of \eqref{clocksignals}. We used the Schwinger-Keldysh formalism \cite{Chen:2017ryl}. Alternatively, one can use the in-in formalism \cite{Weinberg:2005vy,Chen:2010xka,Wang:2013eqj}. 

The quantization of the $X$ field takes the form
\begin{align}
	X_\mathbf k = v_k (\tau) a_{\mathbf k} + v_k^* (\tau) a^\dagger_{-\mathbf k}~.
\end{align}
Then $v_k$ is related to $f_k$ as
\begin{align}\nonumber
	v_k (\tau) & = a^{-3/2} f_k (\tau)   = (-\tau)^{3/2} \sqrt{\frac{\pi}{4}} H e^{-\pi\mu/2} H_{i\mu}^{(1)} (-k\tau) = \alpha_k a^{-3/2} f_k^{\rm out} +\beta_k a^{-3/2} f_k^{\rm out *} ~.
\end{align} 
The second order action for the primordial curvature perturbation is
\begin{align}
S_\zeta = M_p^2 \int d t \frac{d^3 k}{(2\pi)^3} \epsilon (a^3 \dot\zeta^2 - k^2 a \zeta^2)~,
\end{align}
where $\epsilon$ is the slow-roll parameter. Quantizing it in the following way
\begin{align}
\zeta_{\mathbf k} & = u_{ k} c_{\mathbf k} + u_{ k}^* c^\dagger_{-\mathbf k}  ~,
\end{align}
where $c^\dagger_{\mathbf k}$,  $c_{\mathbf k}$ are the creation and annihilation operators satisfying
the usual commutation relations:
\begin{align}
& [c_{\mathbf k}, c^\dagger_{\mathbf p}] = (2\pi)^3 \delta^{(3)} (\mathbf k - \mathbf p)  ~.
\end{align}
The mode function satisfies the following equation of motion
\begin{align}
\ddot u_k +(3+\eta) H \dot u_k +\frac{k^2}{a^2} u_k = 0~.
\end{align}
To the lowest order in slow-roll parameter, the solution is
\begin{align}
u_{k} (\tau) = \frac{H}{2\sqrt{\epsilon} M_{\rm pl}} \frac{1}{k^{3/2} } (1+i k \tau) e^{-i k \tau}~.
\end{align} 

In order to calculate the cosmological collider signal of primordial bispectrum analytically, we first simplify the propagators for $X$, $D_{++}$, $D_{+-}$, $D_{-+}$, and $D_{--}$, as
\begin{align}
D_{++}(\tau_1,\tau_2) = & v_k(\tau_1)v_k^{*}(\tau_2) \Theta(\tau_1 - \tau_2) +
v_k^*(\tau_1)v_k(\tau_2) \Theta(\tau_2-\tau_1)~, \label{eq:D_plus_plus} \\ 
D_{+-}(\tau_1,\tau_2) = & v_k^*(\tau_1)v_k(\tau_2)~,                                 \\
D_{-+}(\tau_1,\tau_2) = & v_k(\tau_1)v_k^{*}(\tau_2)~,                                  \\
D_{--}(\tau_1,\tau_2) = & v_k^*(\tau_1)v_k(\tau_2)\Theta(\tau_1-\tau_2)+
v_k(\tau_1)v_k^{*}(\tau_2) \Theta(\tau_2-\tau_1)~ , \label{eq:D_minus_minus}
\end{align}
where $\Theta$ denotes the Heaviside step function. We can analogously define four types of propagators for $\zeta$, $G_{++}$, $G_{+-}$, $G_{-+}$, $G_{--}$ in terms of  $u_k(\tau)$.

We rewrite the terms in Eq.~\eqref{eq:D_plus_plus} to \eqref{eq:D_minus_minus} in terms of the Bogoliubov coefficients defined in Eq.~\eqref{eq:expand_Bogoliubov}:
\begin{align}\nonumber
v_k(\tau_1) v^*_k (\tau_2) = a^{-3/2} (\tau_1) a^{-3/2} (\tau_2) 
& \bigg[ |\alpha_k|^2 f^{\rm out}_k (\tau_1) f^{\rm out*}_k (\tau_2) + \alpha_k\beta_k^* f^{\rm out}_k (\tau_1) f^{\rm out}_k (\tau_2) \\
& + \alpha^*_k \beta_k f_k^{\rm out*} (\tau_1) f_k^{\rm out*} (\tau_2) +|\beta_k|^2 f_k^{\rm out ^*} (\tau_1) f_k^{\rm out} (\tau_2) \bigg]  ~, \\\nonumber
v^*_k(\tau_1) v_k (\tau_2) = a^{-3/2} (\tau_1) a^{-3/2} (\tau_2) 
& \bigg[ |\alpha_k|^2 f^{\rm out*}_k (\tau_1) f^{\rm out}_k (\tau_2) + \alpha_k\beta_k^* f^{\rm out}_k (\tau_1) f^{\rm out}_k (\tau_2) \\
& + \alpha^*_k \beta_k f_k^{\rm out*} (\tau_1) f_k^{\rm out*} (\tau_2) +|\beta_k|^2 f_k^{\rm out } (\tau_1) f_k^{\rm out *} (\tau_2) \bigg]   ~,
\end{align}
from which we know that only the terms with $|\alpha_k|^2$ and $|\beta_k|^2$ are different. However, the $|\alpha_k|^2$ and $|\beta_k|^2$ terms are local, thus do not contribute to the cosmological collider signal \cite{Arkani-Hamed:2015bza,Lee:2016vti}. In order to understand the bispectrum in the squeeze limit, we focus on the terms proportional to $\alpha_k\beta_k^*$ and $\alpha_k^*\beta_k$, then the four types of propagators $D_{++}$, $D_{+-}$, $D_{-+}$ and $D_{--}$ become identical.
\begin{align}\nonumber
&D(k,\tau_1,\tau_2)\equiv D_{++}(k,\tau_1,\tau_2)=D_{-+}(k,\tau_1,\tau_2)=D_{+-}(k,\tau_1,\tau_2)=D_{--}(k,\tau_1,\tau_2) \\ \nonumber
&= a^{-3/2} (\tau_1) a^{-3/2} (\tau_2) 
 \bigg[ \alpha_k\beta_k^* f^{\rm out}_k (\tau_1) f^{\rm out}_k (\tau_2)  + \alpha^*_k \beta_k f_k^{\rm out*} (\tau_1) f_k^{\rm out*} (\tau_2) \bigg]  ~.
\end{align}
The squeezed-limit bispectrum can be calculated for all orders in the $(k_1/k_3)$ expansion with $k_3\ll k_1 \sim k_2$. Hereafter, we assume $\mu$ is real. The massive scalar propagators become
\begin{align}\nonumber
D(k,\tau_1, \tau_2) &\equiv D_{++}(k,\tau_1,\tau_2)=D_{-+}(k,\tau_1,\tau_2)=D_{+-}(k,\tau_1,\tau_2)=D_{--}(k,\tau_1,\tau_2) \\\nonumber
&=  a^{-3/2} (\tau_1) a^{-3/2} (\tau_2) \bigg[\alpha_k \beta_k^* \bigg(\frac{2H}{k}\bigg)^{i\mu}\bigg(\frac{2H}{k}\bigg)^{i\mu} \frac{\Gamma(1+i\mu)}{\sqrt{2H\mu}}\frac{\Gamma(1+i\mu)}{\sqrt{2H\mu}} J_{i\mu} (-k\tau_1) J_{i\mu} (-k\tau_2) \\
& \quad + \alpha_k^* \beta_k \bigg(\frac{2H}{k}\bigg)^{-i\mu}\bigg(\frac{2H}{k}\bigg)^{-i\mu} \frac{\Gamma(1-i\mu)}{\sqrt{2H\mu}} \frac{\Gamma(1-i\mu)}{\sqrt{2H\mu}} J_{-i\mu} (-k\tau_1) J_{-i\mu} (-k\tau_2)\bigg]
\end{align} 
However, for simplicity, we focus on the first order in $(k_1/k_3)$ expansion, where the Bessel J function is expanded as
\begin{align}
J_{i\mu} (z) = \frac{1}{\Gamma(1+i\mu)} \bigg(\frac{z}{2}\bigg)^{i\mu} ~.
\end{align}
Then the propagator becomes
\begin{align}\nonumber
	D(k,\tau_1,\tau_2) = \frac{a^{-3/2} (\tau_1) a^{-3/2} (\tau_2)}{2H\mu} & \bigg[\alpha_k\beta_k^* \bigg(\frac{2H}{k}\bigg)^{i\mu}\bigg(\frac{2H}{k}\bigg)^{i\mu} \bigg(\frac{-k\tau_1}{2}\bigg)^{i\mu} \bigg(\frac{-k\tau_2}{2}\bigg)^{i\mu} \\
	& +\alpha_k^*\beta_k \bigg(\frac{2H}{k}\bigg)^{-i\mu}\bigg(\frac{2H}{k}\bigg)^{-i\mu}  \bigg(\frac{-k\tau_1}{2}\bigg)^{-i\mu} \bigg(\frac{-k\tau_2}{2}\bigg)^{-i\mu}  \bigg]~.
\end{align}
Noting that
\begin{align}
	\alpha_k \beta^*_k \bigg(\frac{2H}{k}\bigg)^{i\mu}\bigg(\frac{2H}{k}\bigg)^{i\mu}   = \frac{\mu^2\Gamma(-i\mu)^2}{2\pi\mu}~,
\end{align}
and we obtain
\begin{align}\nonumber
	D(k,\tau_1,\tau_2) = \frac{a^{-3/2} (\tau_1) a^{-3/2} (\tau_2)}{4 \pi H }   \bigg[ \Gamma(-i\mu)^2  \bigg(\frac{-k\tau_1}{2}\bigg)^{i\mu} \bigg(\frac{-k\tau_2}{2}\bigg)^{i\mu}  + {\rm c.c} \bigg]~,
\end{align}
and also
\begin{align}
D(p,\tau_1,\tau_2) D(q,\tau_1,\tau_2) = \frac{1}{a(\tau_1)^3a(\tau_2)^3 16\pi^2H^2 } \bigg[\Gamma(-i\mu)^2 \bigg( \frac{ p q \tau_1 \tau_2 }{4}\bigg)^{2i\mu} + {\rm c.c} \bigg]~.
\end{align}
On the other hand, the bispectrum is expressed as
\begin{align}\nonumber
\langle \zeta_{\mathbf k_1} \zeta_{\mathbf k_2}  \zeta_{\mathbf k_3}  \rangle' = 2 c_1 c_2 {\rm Re} \int_{-\infty}^{0}   \int_{-\infty}^{0}   & \frac{d\tau_1}{(-H\tau_1)^3} \frac{d\tau_2}{(-H\tau_2)^2} [\partial_{\tau_1}G_{++}(k_3,\tau_1,0)-\partial_{\tau_1}G_{-+}(k_3,\tau_1,0)]\\
& \int \frac{d^3\mathbf q}{(2\pi)^3} D(p,\tau_1,\tau_2)D(q,\tau_1,\tau_2)
\partial_{\tau_2}G_{++}(k_1,\tau_2,0)\partial_{\tau_2}G_{++}(k_2,\tau_2,0)~.
\end{align}
Since $\mathbf p$ and $\mathbf q$ are constrained by momentum conservation $\mathbf p+\mathbf q - \mathbf k_3=0$,
\begin{align}\nonumber
& \langle \zeta_{\mathbf k_1} \zeta_{\mathbf k_2}  \zeta_{\mathbf k_3}  \rangle' =  2 c_1 c_2 {\rm Re} \int_{-\infty}^{0}   \int_{-\infty}^{0}    \frac{d\tau_1}{(-H\tau_1)^3} \frac{d\tau_2}{(-H\tau_2)^2} [\partial_{\tau_1}G_{++}(k_3,\tau_1,0)-\partial_{\tau_1}G_{-+}(k_3,\tau_1,0)]\\
& \quad \int \frac{d^3\mathbf q}{(2\pi)^3}\int \frac{d^3\mathbf p}{(2\pi)^3} (2\pi)^3 \delta^{(3)}(\mathbf p+\mathbf q- \mathbf k_3) D(p,\tau_1,\tau_2)D(q,\tau_1,\tau_2)
\partial_{\tau_2}G_{++}(k_1,\tau_2,0)\partial_{\tau_2}G_{++}(k_2,\tau_2,0)~.
\end{align}
First we write the delta function $\delta^{(3)}(\mathbf p+\mathbf q-\mathbf k_3)$ into an integration of exponential function in $\mathbf x$ space and integrate out $\mathbf q$ and $\mathbf p$ by Fourier transform. To write this procedure more explicitly, we have
\begin{align}
F(k_3) & = \int \frac{d^3 \mathbf p}{(2\pi)^3} \int \frac{d^3 \mathbf q}{(2\pi)^3} (2\pi)^3 \delta^{(3)} (\mathbf p + \mathbf q - \mathbf k_3) p^m q^n \\
&  = \int d \mathbf x \int \frac{d^3 \mathbf p}{(2\pi)^3} \int \frac{d^3 \mathbf q}{(2\pi)^3}  e^{-i (\mathbf p + \mathbf q - \mathbf k_3) \cdot \mathbf x } p^m q^n~,
\end{align} 
and the Fourier transform is given by
\begin{align}
\int d^3 \mathbf x e^{i \mathbf k\cdot \mathbf x} x^{-\Delta} & = k^{\Delta-3}~, \\
\int \frac{d^3 \mathbf p}{(2\pi)^3} e^{-i \mathbf p\cdot \mathbf x} p^{\Delta} & = x^{-\Delta-3}~.
\end{align}
Then, we integrate out $\mathbf x$. After this procedure, we obtain
\begin{align}\nonumber
& \langle \zeta_{\mathbf k_1} \zeta_{\mathbf k_2}  \zeta_{\mathbf k_3}  \rangle' =  2 c_1 c_2 u_{k_1}(0)u_{k_2}(0)u_{k_3}(0) {\rm Re} \int_{-\infty}^{0}   \int_{-\infty}^{0}    \frac{d\tau_1}{(-H\tau_1)^3} \frac{d\tau_2}{(-H\tau_2)^2} [\partial_{\tau_1}u_{k_3}(\tau_1)-\partial_{\tau_1}u^*_{k_3}(\tau_1)]\\
& \quad   \frac{1}{a(\tau_1)^3a(\tau_2)^3 16 \pi^2 H^2 } \bigg[ \Gamma(-i\mu)^4 \bigg( \frac{ k_3^2 \tau_1 \tau_2 }{4}\bigg)^{2i\mu} + {\rm c.c} \bigg] \partial_{\tau_2}u_{k_1}(\tau_2)\partial_{\tau_2}u_{k_2}(\tau_2)~,
\end{align}
where we have used the fact that $u_k(0)=u^*_k(0)$. Finally, we obtain the non-Gaussianity as
\begin{align}\label{clocksignal}
\langle \zeta_{\mathbf k_1}\zeta_{\mathbf k_2}\zeta_{\mathbf k_3} \rangle' = c_1 c_2 {\rm Re} \bigg[ g(\mu) \frac{2^{-1-8\mu}H^5}{k_1 k_2(k_1+k_2)^4 M_{\rm pl}^6 \epsilon^3 } \bigg(\frac{k_1+k_2}{k_3}\bigg)^{2i\mu} \bigg]~,
\end{align}
where the factor $g(\mu)$ is given by
\begin{align}
g(\mu) = \frac{\Gamma(2-2i\mu)\Gamma(4-4i\mu)\Gamma(-2i\mu)^4}{\Gamma(1/2-i\mu)^2 \Gamma(1/2+i\mu)^2} \sinh^2(\pi\mu)~.
\end{align} 
%For three external legs, see \cite{Barnaby:2011vw} for numerical treatment of the loop integration.

% Generated by or.py

\end{document}